\documentclass[12pt]{article}
\usepackage{psfrag}
\renewcommand{\d}{\dagger}
\newcommand{\oh}{ \frac{1}{2} }

\newcommand{\niu} {n_{i\uparrow}}

\newcommand{\nid} {n_{i\downarrow}}

\newcommand{\s} {\sigma}
\newcommand{\aid}{a_{i \s}^{\dagger} }
\newcommand{\ai}{a_{i \s} }
\newcommand{\la}{\langle}
\newcommand{\ra}{\rangle}

\newcommand{\aiu}{a_{i \uparrow}}

\newcommand{\aidd}{a_{i \downarrow}}

\newcommand{\aj}{a_{i \s} }
\newcommand{\be}{\begin{equation}}
\newcommand{\ee}{\end{equation}}
\newcommand{\beq}{\begin{eqnarray}}
\newcommand{\eeq}{\end{eqnarray}}
\newcommand{\eps}{\epsilon}
\newcommand{\nt}{N_t}
\newcounter{x}
\newcommand{\PRL}{ Phys.\ Rev.\ Lett.\ }
\newcommand{\PR}{  Phys.\ Rev.\  }
\newcommand{\fc}{Fig.\ \arabic{x} \addtocounter{x}{1}}
\begin{document}
\vspace{3cm}
\begin{center}
{\large \bf  Hubbard Model with L\"uscher fermions} \footnote{Work supported
in part by the KBN grants: no.~2P03B19609 and no.~2P03B04412}
\end{center}
\vspace{1.0cm}

\begin{center}

{\sl Pawe\l \ Sawicki } \footnote{A fellow of the Foundation for Polish
Science}
\vspace{0.5cm}

{\sl Institute of Physics, Jagiellonian University, \\
ul. Reymonta 4, PL-30-059, Krak\'ow }
\end{center}

\vspace{1.5cm}
\begin{center}
{\bf Abstract }
\end{center}
We study the basic features of the two-dimensional quantum Hubbard Model at 
half-filling by means of the L\"uscher algorithm and the algorithm based 
on direct update of the determinant of the fermionic matrix. We implement the
L\"uscher idea employing the transfer matrix formalism which allows
to formulate the problem on the lattice in $(2+1)$ dimensions. We discuss
the numerical complexity of the L\"uscher technique, systematic errors
introduced by polynomial approximation and introduce some improvements
which reduce long autocorrelations. In particular we show that  
preconditioning of the fermionic matrix speeds up the algorithm and
extends the available range of parameters. We investigate the
magnetic and the one-particle properties of the Hubbard Model at
half-filling and show that they are in qualitative agreement with the
existing Monte Carlo data and the mean-field predictions.

\vspace{2.9cm}
\begin{tabbing}
\mbox{} \= TPJU--4/97  \\
\mbox{} \= March 1997 \\
\end{tabbing}
\newpage
\tableofcontents
\newpage

\section{Introduction}

It is very well known that quark degrees of freedom are difficult
to include in numerical simulations of quantum systems \cite{qg}.
Although the substantial progress has been made in this field
the search for new algorithmic techniques is an active
research area. At present probably the best method is to integrate
the quark degrees of freedom and simulate the theory with effective
action. However such action usually contains determinant which is
a nonlocal object and its evaluation is very expensive in computer
time. Recently a new computational scheme has been proposed by
L\"uscher who showed, in the QCD case, how to approximate determinant
of the Dirac operator by a local bosonic action \cite{lu}. So far 
there is no  complete discussion of properties of this algorithm and
it is of great interest to perform some tests of its performance.

Besides QCD there is a wide class of physically interesting problems
where effective technique for simulating fermionic degrees of freedom
is of crucial importance. The Hubbard Model \cite{hub} is an example of
a conceptually simple model with effective Coulomb repulsion between 
electrons which for many years has been the basic framework for studying
strongly correlated electrons in solid state physics.  The interest of 
the Hubbard Model has been renewed after finding that it can qualitatively 
describe the magnetic properties of the high $T_c$ superconducting 
materials above the temperature of the superconducting transition \cite{dag}.
There is also a conjecture suggested by a variety of calculations that
when two dimensional system is doped slightly away from half-filling the
repulsion between electrons can give rise to superconductivity \cite{and}.
Until now this hypothesis has not been confirmed or rejected.
The Hubbard Model is very well suited to tests of the new fermionic
algorithms because it contains all details of fermionic formulation and
describes very interesting, and rather unexplored from numerical point 
of view, physics of strongly correlated electrons.

In this paper we apply the L\"uscher idea to the Hubbard Model. 
We show how to study basic features of strongly correlated electrons 
and note about the problems which can be solved by numerical simulations. 
We study the efficiency of the L\"uscher algorithm in comparison with 
the algorithm based on direct update of the determinant of the fermionic 
matrix. We propose also some improvements which reduce long autocorrelation 
times. Some conclusions from this dicussion can be applied in 
implementation of  the L\"uscher algorithm in other areas.

The paper is organized as follows.  Section 2 is assigned to the review
of the basic properties of the Hubbard Model. We emphasize some
important not fully solved problems. Section 3 covers path integral
formulation of the Hubbard Model.  In section 4 we review fermion
Quantum Monte Carlo methods and  introduce two new algorithms. One is
the implementation of the L\"uscher
method to the Hubbard Model. The second is a modification of the algorithm
based on direct update of the determinant of the fermionic matrix.
In section 5 we compare dynamical properties of these two algorithms and
present some physical results obtained from simulations of half-filled
band Hubbard Model. The comparison of these results with
previous findings and theoretical predictions gives credence to our
methods of simulations. Section 6 contains a summary and our
conclusions.

\section{The Hubbard Model}
The  Hubbard Model is defined by the Hubbard Hamiltonian
\be
{\cal H}= - K \sum_{\la i,j \ra, \s} \aid {a_{j \s} } +U \sum_{i}
\left(   n_{i \uparrow} -\oh  \right)
\left(   n_{i \downarrow} -\oh \right)
-\mu \sum_{i \s} n_{i \s}  \label{one} ,
\ee
where $ a_{ i \s } $ is the annihilation operator of an electron
with spin $\s$ located at site $i$ of $d$ dimensional lattice.
In this paper however we will concentrate  mainly on the
case $d=2$, which is physically the most interesting.
The summation in the kinetic term runs over nearest neighbors
(each symmetric pair of $\la i,j \ra$ is
counted twice) and the hopping energy $K$ can be taken as a unit of energy.
The interaction term takes into account short-range effective
Coulomb repulsion. The term with the chemical
potential determines the band filling. The half--filled band has $\mu=0$.
There are three dimensionless parameters which determine the behaviour
of the system, interaction energy $U/K$, band filling $\mu/K$ and
temperature $T/K$. In this parameter space the Hubbard Model exhibits various
phases.

Originally the Hubbard Model was introduced as the simplest system
which might exhibit an insulating (Mott) state. It is exactly soluble only
in one dimension. In this case the integral equation for the density of
states can be solved in a closed form \cite{wu}. It follows that the
system is insulating for any nonzero $U$ and conducting for $U=0$. Thus
there is no Mott transition at finite $U$. However even for
1d system we have only partial information. For example, the correlation
functions are not known.
In 2 and higher dimensions many different approximate techniques have
been used. Mean field approximation \cite{fra}, variational techniques
\cite{varia} and other were successful in describing the properties of
strongly
correlated electrons. They are very important because they give insight
into physics. Unfortunately  it is
very difficult to judge the accuracy of approximate techniques beyond
the perturbative limit.
Therefore it is not surprising that they often give conflicting results.
Numerical simulations can fill this gap and deliver nonperturbative results
for comparison.

Some properties of the Hubbard Model on square lattice may be understood
from simple qualitative considerations. Let us start from
the noninteracting limit. The dispersion relation
\be
\eps_k = -2 K (\cos{k_x} + \cos{k_y} ) , \label{disp}
\ee
establishes the band structure. The total bandwidth
$E_b=8K$. The ground state
is obtained by populating the states from the bottom of the energy scale
up to the Fermi level. In the half-filled case the rectangular shape of
the Fermi surface is determined from the condition $\eps_F =0 $
as shown in  \fc.

The existence of parallel sectors on the  Fermi
surface which are separated by the vector $Q_{nest}=(\pm \pi, \pi)$
(nesting) has important physical consequences and
is crucial for the developing of the magnetic instabilities. Indeed
the weak interaction could induce the process of exchanging electrons
with opposite spins from opposite sides of the Fermi surface. It
involves however the exchange of momentum $Q_{nest}$  which is
characteristic for antiferromagnetism.  Obviously without detailed
theory it is impossible to  make final conclusions. For example, there
is also a process of quasi-forward scattering with the exchange of
momentum $(0,0)$ which would give rise  to ferromagnetism and from
simple qualitative arguments it is impossible to judge which is
dominant.

On the other hand, in the strong coupling limit $(U \rightarrow
\infty)$ the interaction term forces electrons not to reside on the
doubly occupied site.
The hopping term can be treated as a perturbation and the ordinary
expansion in the parameter $K/U$ is applicable.
The effective Hamiltonian is the Quantum Heisenberg Antiferromagnet
\be
{\cal H}_{0} = \frac{2 K^2}{U} \sum_{ \la i,j \ra} \vec{S_i} \vec{S_j},
\ee
with the exchange coupling $2 K^2 /U$. The above result is valid for
the half-filled system in any dimension and lattice.

Therefore it is plausible that the Hubbard Model may
possess ground state with nontrivial magnetic properties. More
quantitative description  can be obtained by the mean field theory.
One assumes that the ground state has some sort
of magnetic order and the problem can be solved by means of variational
principle. Detailed calculations for the antiferromagnetic ground state
is given in the Appendix A. Here we stress main results.
One possible solution is the {\em ferromagnetic} state which can
exists for sufficiently large $U$. The critical value is given by
Stoner's criterion
\be
U_{crit} \rho( \eps_F ) >1
\ee
with $\rho(\eps_F )= \sum_k \delta( \eps_F -\eps_k) $ being the density
of states on the Fermi surface. But there is also an {\em
antiferromagnetic} solution for any $U$
with the smaller energy which is a better candidate for the true ground
state of the half-filled Hubbard Model. Numerical simulations fully
confirm that expectation and the antiferromagnetic order has been
explicitly proven for $U \leq 2$ \cite{anti}. It is also believed that
antiferromagnetism exists for any nonzero $U$.
Mean field solution gives also new band structure $E_k= \pm
\sqrt{\eps_{k}^{2} + \Delta^2} $ with the gap $2 \Delta$ above the Fermi
surface characteristic for an insulator.

A lot of single-particle properties can be obtained from the study of
the one particle Green's function defined by
\be
G_{\s \s '}(r,t; r',t')= -i \la  T a_{\s}(r,t) a_{\s '}^{\dagger} (r',t') \ra
\ee
where $a_{\s}(r,t)$ are annihilation operators in Heisenberg picture. $T$
orders  product of operators at earlier time to the right of operators
at later time, keeping track of the sign changes associated
with Fermi statistic;
i.\ e.\ for two fermionic operators $A(t)$ and $B(t)$ we have
\be
TA(t)B(t')= A(t) B(t') \theta (t -t') - B(t') A(t) \theta (t' -t) .
\ee

For the noninteracting system, which is translationally invariant
and time independent the Fourier transform of $G(r,t; r',t')$
is a function of 2 variables \cite{fet}
\be
G_{\s \s'}^{0} (k, \omega)=\delta_{\s \s'} \lim_{\nu \rightarrow 0+}
\left( \frac{\theta(\eps_k - \eps_F )}{\omega -\eps_k +i \nu}
+  \frac{\theta(\eps_F - \eps_k )}{\omega -\eps_k -i \nu} \right).
\ee
The poles of the propagator exhibit typical one particle spectrum.
In the weakly interacting case we can expect that physical low-energy
excitations look like weakly interacting fermions.
The propagator retains its pole structure with a renormalized dispersion
relation $\eps_{ren}(k)$.  Thus it is expected that the propagator near
the Fermi surface  should look like
\be
\lim_{\omega \rightarrow \eps_F}    G(k, \omega)  =
\lim_{\nu \rightarrow 0+}  \left( \frac{Z}{\omega -\eps_{ren}(k) +i \nu}
             +G_{reg} \right)
\ee
with a certain regular part $G_{reg}$. The {\em wave function
renormalization Z} measures the discontinuity of the occupation
number $n_{k \s}= iG_{\s \s} (k,t=0^{-}) $ at the Fermi surface.
The system which exhibits such behavior is called a {\em Fermi liquid}.

The ground state of the half-filled Hubbard Model has long range
antiferromagnetic order being the result of strong interactions. Thus we
may expect quite different behaviour than the mentioned above. In fact
the renormalization constant $Z$ vanish and simulations fully confirm
the lack of sharp Fermi surface \cite{white} \cite{scalapino}.

What happens after doping still remains unclear.
The approximate analytic methods are not conclusive and one has to rely
on numerical calculations. Unfortunately in this case the fermionic
determinants are not positive definite. As will be discussed
in Section 4 this difficulty called the sign problem significantly
restricts the range of predictions. Nevertheless quantum MC can still
deliver useful informations. For example, it was 
confirmed that antiferromagnetic order is destroyed even for small
doping \cite{white} \cite{scalapino}. At the same time however it 
is not certain weather the
ground state becomes  the Fermi liquid. The simulations performed at the
quarter filling strongly suggest the existence of the sharp Fermi surface.
The situation is unclear at small doping where the sign problem
is very severe. Recent simulations \cite{sorella} suggest the
vanishing $Z$ as the size of lattice increases at $U=4$. However in the last
calculations the small magnetic field was introduced to alleviate
the sign problem and it is not clear how it can influence results.
Thus more work should be done to clarify this issues.
Should the non Fermi liquid behaviour be
confirmed the holes would not propagate as free entities. One possible
scenario is binding of holes and possible existence of superconducting phase.

The problem of the existence of the superconducting state has been studied 
directly in numerical simulations \cite{white} \cite{scalapino}. 
The appropriate observable is the correlation function for pairs
\be
P(r) = \la \Delta^{\dagger}_i \Delta_{i+r} \ra .
\ee
The annihilation operator for pair is defined as
\be
\Delta_i = a_{i \uparrow} (a_{i+ {\hat x}  \downarrow} +
                           a_{i- {\hat x}  \downarrow}
               \pm a_{i+ {\hat y}  \downarrow} \pm  a_{i+ {\hat y}
 \downarrow} )
\ee
where the $(+)$ sign correspond to extended $s^\star$ wave, and
the $(-)$ sign to $d_{x^2 -y^2}$ wave, as can be seen from rotational
symmetry \cite{scal95}. In the superconducting state these correlations
should diverge with the spatial lattice volume. A widely used quantity
to monitor such behaviour is the susceptibility
\be
\chi = \sum_r P(r).
\ee
Current simulations indicate only very weak dependence of the pair
correlations  on lattice size what does not confirm the
existence of the superconducting phase. One may argue however  that the
lattice sizes available for computation are too small while comparing
with the correlation length for the pair.

The Hubbard Model can be modified in many ways. Adding the next-nearest
neighbour hopping has very important consequences because the Fermi
surface ceases to be nested. In this case critical value of $U$ exists
for the appearance of ferromagnetism. There is also some evidence that
such model would be more relevant to the problem of high $T_c$
superconductivity and recent numerical results indicate enhancement of
correlation functions for pairs \cite{santos}. These issues are still
intensively studied.

Contrary to the results discussed above the superconducting phase
has been found for the attractive Hubbard Model i.e. with the reversed sign
in front of $U$ \cite{negative}. In this case the QMC simulations
are much simpler since the sign problem does not  occur.
It might appear  that physics of the attractive and the repulsive model
should be very similar because both  models
are related by a change of sign in $U$ and the exchange of chemical
potential and magnetic field. However there is no evidence that
the attractive model can serve us as an effective model for some range
of parameters of the repulsive model, where one would expect pairing.
Thus, at present simulations of the attractive model simply are not
a solution of the sign problem.

\section{Path integral formulation of the Hubbard Model}
In this section we derive the path integral formulation of the Hubbard Model.
Decoupling of interaction terms by means of the Hubbard-Stratonovich
transformation and applying transfer matrix formalism allows
to write the expression for the partition function in such a form that it
can be used as a starting point for L\"uscher local bosonic approximation.

Our derivation  of the Euclidean path integral partition function
follows closely that presented by Creutz \cite{creutz}. All
arguments apply essentially for both: the attractive and the repulsive Hubbard
Model. However there are also some important differences which  will be
stressed. We begin with the Hamiltonian Eq. (\ref{one}), which may
be rewritten as
\beq
{\cal H}= -K \sum_{ \la i,j \ra \s} \aid a_{j \s}-\frac{U}{2}
\sum_{i} (\niu - \nid)^2 -\mu \sum_{i \s} n_{i \s}.
\eeq
We assume that $i$ and $j$ are sites on a two dimensional square lattice
with $N=N_{x}^{2}$ sites.
This formula differs from the standard one for the repulsive Hubbard
Model by an irrelevant additive constant.

All thermodynamic functions for  a many particle system in the
temperature $T$ can be obtained from the partition
function
\be
Z= Tr [ \exp{- ( {\cal H} /T } ) ].
\ee
$Tr$ denotes  the sum over the complete set of physical states.  The thermal
expectation   value of some physical observable, say ${\cal O}$, is
defined by
\be
 \la {\cal O} \ra = Tr [ {\cal O}
 \exp{- ( {\cal H} /T } ) ]/Z
 \label{trasa}.
\ee
Unfortunately, it is  usually very difficult to evaluate the trace
in Eq. (\ref{trasa}) exactly for physically interesting systems.
Lattice field theory offers an interesting possibility to evaluate
the expectation values for finite systems employing stochastic methods.

In the lattice  formulation one interprets  the inverse temperature,
$1/T$, (which is also referred to as Euclidean time) as the extra
dimension and discretizes it dividing the interval $(0,\beta=1/T)$
into $N_t$
slices of length $\eps=\beta/\nt$. The partition function can be
written as
\be
Z=Tr[{(e^{ - \beta {\cal H} /\nt})}^{\nt} ].
\ee
In the next step the separation between kinetic ${\cal K}$ and interaction
${\cal V}$ term is performed according to the Trotter formula
\be
e^{\eps {\cal H}}=e^{\eps {\cal K} } e^{\eps {\cal V} } +O({\eps}^2),
\ee
which starts the series of approximations in which the terms of order
of $O( (\beta/\nt)^2 ) $ are consistently neglected
\beq
e^{-\beta {\cal H} /\nt} & = & \exp \left[ \frac{K \beta}{\nt}
  \sum_{  \la i,j \ra \s}
\aid \aj   \right] \nonumber \\
 & & \exp{ \left[ \frac{U \beta}{2 \nt} \sum_{i} {(a_{i \uparrow}^{\dagger}
 \aiu - a_{i \downarrow}^{\dagger} \aidd)}^2
 +\frac{\mu \beta} \nt \sum_{i, \s} \aid \ai \right] } .
\eeq
As all the terms presented in the last exponent commute with each other
we can treat them as ordinary numbers. In particular one can
use the identity
\be
\int_{-\infty}^{\infty} e^{-ax^2/2 + bx}\, dx =
{(2 \pi)}^{1/2}e^{b^2/{2a} },\; \; a>0 \; ,
\ee
which is a continuous version of the Hubbard-Stratonovich
transformation. With this in mind we can introduce the integration over
auxiliary set of variables ${A_i}$ located on the lattice sites to
decouple the quartic term
\beq
\lefteqn{     e^{-\beta {\cal H}/ \nt}  = (2 \pi)^{-N^2/2} \int [ DA_{i} ]
e^{-\sum_i A_{i}^2/2} \exp \left[ \frac{K \beta}{\nt}  \sum_{  \la i,j \ra \s}
\aid \aj   \right]   } \nonumber \\
& & \exp \left[ \sum_i (U \beta/\nt)^{1/2} A_i
(a_{i \uparrow}^{\dagger} \aiu - a_{i \downarrow}^{\dagger} \aidd)
+\frac{\mu \beta}{\nt} \sum_{ i,\s} \aid \aj \right]  .
\eeq

Note an ambiguity which appears in the last expression, namely which sign
should have square root of ${(n_{i \uparrow} - n_{i \downarrow})}^2$.
Two possible options are equally good and we arbitrary choose one. In
principle the asymmetry between up and down components should vanish
after performing functional integral over $A_{i}$ and should not
affect the measured observables. However due to further finite $\nt$
approximations the  corresponding results for two spin components
will differ in real simulations.

To clarify the positivity of the final fermionic determinant it is
convenient to formally recover the symmetry between up and down
polarizations by performing the particle-hole
transformation. Introducing  the creation operators for hole
$ b_{i}^{\dagger}=(-1)^{i_x + i_y} a_{i \downarrow}$ and denoting
$a_{i \uparrow}$ simply by $a_i$ we get
\beq
\lefteqn{  e^{-\beta {\cal H} /\nt}  = (2 \pi)^{-N^2/2} \int [ DA_{i} ]
e^{-\sum A_{i}^{2}/2}
 \exp { \left[ \frac{K \beta}{\nt}  \sum_{  \la i,j \ra }
( a_{i}^{\dagger} a_j  +b_{i}^{\dagger} b_j)    \right]  }  }\nonumber \\
&  & \exp{ \left[ \sum_i (U \beta/\nt)^{1/2} A_i ( a_{i}^{\d} a_i +
b_{i}^{\dagger} b_i -1)
-\frac{\mu \beta}{\nt} \sum_i ( a_{i}^{\d} a_i + b_{i}^{\d} b_i) \right] }
,
\eeq
which is explicitly symmetric in $a_i$ and $b_i$ operators.
The sum in the exponents can be expressed as products with help
of the following relations
\beq
\exp{ (a_{i}^{\dagger} a_j )} = 1 +  a_{i}^{\dagger} a_j \; \; ( i \neq j)
& , \; \; \; \;  \exp{( \alpha a_{i}^{\dagger} a_i )} =
1 +  a_{i}^{\dagger} a_i (e^\alpha -1)
\eeq
After collecting time slices together and shifting the $A_i$ integration
the resulting trace of normal ordered products takes the form:
\beq
\lefteqn{ Z=
\exp{ [N^2 \beta( U/2)+ \mu)]{(2 \pi)}^{-N^2 \nt/2 } }   } \nonumber \\
&  \int [D A_{i,t}] e^{-\sum_{i,t} A_{i,t}^2/2} Tr \prod_t :
\left[ \prod_{ \la i,j \ra} [1 + (K \beta/ \nt) a_{i}^{\d} a_j] \nonumber \right. \\
& \prod_i (1 + a_{i}^{\d} a_i \{ \exp{ [{(U \beta/ \nt)}^{1/2} A_{i,t}
-(U - \mu) \beta
/ \nt] -1 }  \}  )  \nonumber \\
 & \left. (\{ a, a^\dagger, \mu \} \rightarrow \{ b, b^{\dagger}, -\mu \})
 \right]:
\eeq
For every time slice, labeled by a discrete index $t=1,\ldots , \nt$ the
anticommutation relations can  be used to convert  this expression into
a sum of normal ordered operators,
neglecting higher orders in $\beta/\nt$ . The trace can be expressed as a
Grassmann integral. In Appendix B we present the general formalism  of
the Grassmann variables and the derivation of the partition function.
The final result is
\beq
Z= {\cal N} \int [DA_{i,t}] e^{ - \sum_{i,t} A_{i,t}^{2}/2 }
 \det{M_+} \det{M_-}, \label{grassman}
\eeq
where ${\cal N}$ is some normalization constant.
The matrices $M_+$ and $M_-$ are specified by their elements
between arbitrary vectors $\psi^\star $ and $\psi$
\beq
 \psi^\star M_\pm \psi & = &\frac{K \beta}{\nt} \sum_{ \la i,j \ra,t}
\psi_{i,t}^{\star} \psi_{j,t} + \sum_{i,t} \psi^{\star}_{i,t} (\psi_{i,t}
- \psi_{i,t-1}) \nonumber \\
& & +\sum_{i,t} \psi^{\star}_{i,t} \psi_{i,t}
\{ \exp{[(U \beta/\nt)^{1/2} A_{i,t}
-(U \mp \mu) \beta / \nt] -1}  \}. \label{matrix}
\eeq
The antiperiodic boundary conditions  $\psi_{i,0}= - \psi_{i,N_t}$ are
taken in the $t$ direction and reflect the Grassmann nature of the
fermionic fields (see Appendix B). The $M$ matrices are not Hermitian.
It seems to be an intrinsic property of the systems with Galilean group
of symmetry and can not be simply avoided \cite{creutz1}. This fact is rather
disadvantageous for effective implementation of the L\"uscher method as
can be seen later.

Our expression for the partition function is rather nonstandard. We decided
to work in the representation with  dimension $V=N \nt$ to avoid
multiplication  of matrices present in the standard formulation (cf Eq.
(\ref{h})). Moreover to obtain the simple
structure for the matrix $M$ we performed the series of approximations
in the parameter $\eps = \beta /\nt$  treating the higher order terms
in such a way to get the simplest possible form of matrices $M_{\pm}$.
Due to these approximation our expression for the partition function
becomes exact only as $\nt \rightarrow \infty$. In principle all
measured quantities should be obtained by extrapolation to this limit.

The half-filled case is favoured from the computational
point of view because then the statistical weights are positive
\be
\det{M_{+}} \det{M_{-}}=\det{M}^2
\ee
since $M=M_-=M_+$.
The same it is not true outside half-filling and the sign problem
occurs (see Section 4).  Attractive Hubbard model is more tractable as the
distribution is positive definite for any $\mu$. The interaction term can
be written as
\be
-\frac{|U|}{2} \sum_{i} (\niu + \nid)^2.
\ee
and the derivation follows essentially the same steps. However in
contrast to the previous case after performing the Hubbard-Stratonovich
transformation the transfer matrix remains symmetric in the spin
components  and the particle hole transformation is superfluous.
The statistical weights in the final result are evidently positive definite
\beq
Z= {\cal N'} \int [DA_{i,t}] e^{ -\sum_{i,t} A_{i,t}^{2}/2 }
 (\det{M_+})^{2}.
\eeq
The matrix $M_+$ is still given by Eq. (\ref{matrix}) with $U$
replaced by $|U|$.

\section{ Algorithms }

\subsection{Introduction}

The numerical difficulties with fermionic fields come from their
anticommuting nature. The transfer matrix is an operator in a Grassmann
space and its matrix elements cannot be interpreted directly as
a probability  in Monte Carlo algorithms. For many interesting problems
the formalism similar to the described above can be used to integrate
the fermionic  degrees of freedom.  Once the final expression for the
partition       function is given in closed  numerical form standard
techniques of stochastic sampling can be applied.

Unfortunately fermionic determinants appearing in the partition
function are non--local objects. This means that local updates require
calculations  which depend on the state of the whole system. Thus to
update the fields  individually on each lattice site, or link, soon
becomes prohibitively expensive in computer time as the size of system
grows. At present, for example in the QCD case, the best known algorithms
update many degrees of freedom simultaneously.

Additionally to the nonlocality of the action one has often another
difficulty with lattice fermions known as a doubler problem. It is
for example evident in the lattice formulation of QCD where the species
of quarks are doubled in the simplest scheme incorporating the chiral
symmetry. There are two popular approaches for dealing with this
difficulty. In the Kogut-Suskind formulation each site carries only a
single component of the Dirac spinor. In the Wilson
formulation chiral symmetry of Lagrangian is explicitly broken
and it is expected to be restored only in the continuum limit
\cite{wosiek}.
The Hubbard Model is physically defined on the lattice and hence the
doubler problem does not appear in its path integral formulation.

In this section we review some fermionic algorithms. We discuss the
algorithms based on evolution equations which are widely used in
QCD and algorithms developed specially for simulations of the
Hubbard Model. Afterwards, we present two particular algorithms which we
use for simulations of the Hubbard Model. One is the application of
the  L\"uscher idea to the path integral representation, which we
discussed in the previous section. The second is a modification
of the algorithm based on direct computation of the determinant
and was introduced as a reference point  for the
comparison of results and efficiency of the L\"uscher algorithm.

\subsection{Algorithms based on evolution equations}
To be more specific we concentrate  on the action typical
for fermionic problems
\be
Z=\int [DA] e^{-S_B (A) } \det {\cal M}(A) \label{xx}.
\ee
We denote the bosonic part of the action depending on the field $A$
(which is the Hubbard--Stratonovich field for the Hubbard Model)
by $S_B (A) $. The matrix ${\cal M}$ contains all details of
fermionic formulation and its explicit form depends on the model.
We assume that matrix ${\cal M}$ is positive definite.
For example in simulations of the half-filled Hubbard Model
${\cal M}=M^{\dagger} M $ with matrix $M$ given by Eq.
(\ref{matrix}).

The determinant of the matrix ${\cal M}$ can be expressed as a gaussian
integral over a set of auxiliary complex fields $\phi$
\be
Z=\int [DA] [d \phi] [d \phi^{\star}] e^{ -S_B + \phi^{\star}
{\cal M}^{-1}
\phi}.
\label{zz}
\ee
The problem of evaluating determinant is now reduced to the of inversion
of matrix ${\cal M}$. Each time components of the $A$ field are
changed the evaluation of new ${\cal M}^{-1}$ matrix is required. The
problem is slightly simplified because for local actions ${\cal M}$ is
sparse. In this case there are some useful iterative schemes appropriate
for this task. Commonly used is the conjugate gradient algorithm \cite{hmc}.

However the inversion of the matrix still requires a lot of CPU
time. Hence to reduce the computational effort many approximate
algorithms have been proposed. One of the simplest and interesting is
the {\em pseudofermion} method \cite{marinari}. In this approach one
considers only small changes in the $A$ field linearizing the change
of the action
\be
\frac{dS}{dA} = \frac{dS_B}{dA} -Tr \left( {\cal M}^{-1}
\frac{d {\cal M} }{dA}
\right)
\ee
with
\be
S= S_B -Tr \log{ {\cal M}(A)} .
\ee
For the local matrix ${\cal M}$ the quantity $\frac{d {\cal M} }{dA}$
can be
easily calculated. The elements of the ${\cal M}^{-1}$ can be obtained
as the appropriate correlation functions
\be
({\cal M}^{-1})_{i j} = \la \chi^{\star}_j \chi_i \ra .
\ee
The expectation value is taken over complex fields $\chi$, called
{\em pseudofermions}  which are distributed according to the formula
\be
P(\chi) \propto \exp{(-\chi^\star {\cal M} \chi) }.
\ee
Fields $\chi$ can be easily simulated giving the estimation for the
elements of matrix  ${\cal M}^{-1}$. Such estimation is usually done
once per full sweep of the variables $A$. The assumption of small changes,
which can be realized by taking sufficiently small step for proposing
trial values of the $A$ field is very important. The algorithm is only
approximate and in  principle  all measured  quantities should be
obtained by extrapolation   with the size of the step going to zero. In
practice such analysis is  rarely made due to the insufficient computing
resources. The algorithm described above usually has long
autocorrelations and is inefficient. However it was the starting point
for developing new better approaches.

Another very  friutfull idea refers to stochastic equations, which can
be used to generate the fields with the probability distribution
(\ref{xx}). For sake of simplicity we restrict the discussion to a
single degree of freedom. Further generalization to the field theory is
straightforward. Let us consider a particle with mass $m$ moving in a
potential $V(x)$.  To the Newton's equation  of motion we add also a
drag force proportional to the velocity and a randomizing force
\be
m \frac{d^2 x}{d \tau}= -\frac{dV}{dx} - \alpha \frac{dx}{d \tau}
                        +\left( \frac{2 \alpha}{\gamma} \right) ^{1/2}
                         \eta (\tau)    \label{evol}
\ee
with $\alpha$ and $\gamma$ being free parameters. The random force is
white noise
$\la \eta (\tau) \eta(\tau ') = \delta(\tau- \tau ') \ra $.
Detailed definition of the distribution $\rho (\eta)$ of will be given
later.  The coefficient in front of $\eta$ is a matter of convention.
After introducing the momentum $p$ the equation can be rewritten as a
system of two first order equations
\beq
 \frac{dp}{d \tau} &= &-\frac{dV}{dx} - \frac{\alpha p}{m}
         +\left( \frac{2 \alpha}{\gamma} \right) ^{1/2} \eta (\tau)
                    \nonumber \\
            \frac{dx}{d \tau}&=& \frac{p}{m}  .
\eeq
For the simulation purposes we divide the time $\tau$ into slices
of size $\eps$. The quantities $x'$ and $p'$ at time $\tau+\eps$ can
be calculated from $x$ and $p$ at time $\tau$ from the
formulas
\beq
 p'& =& p + \eps \left[ -\frac{d V}{dx} - \frac{ \alpha p}{m}
         +\left( \frac{2 \alpha}{\gamma} \right) ^{1/2} \eta (\tau)
    \right] , \nonumber \\
     x'&=& x + \frac{\eps p'}{m}.
\eeq
To minimize the errors of discretization special form of update  was
introduced. The momentum $p$ is calculated first and this new value
is used to update $x$.
In the limit of deterministic evolution this "leap-frog" scheme of
integration  preserves phase space volumes ( the Jacobian of the
transformation equals 1)  and is reversible under the change of the
sign of $\eps$ . These properties are very useful in making the
algorithm exact.

The quantity $\eta$ is generated, independently for each discretized
slice of the evolution time, according to distribution
\be
\rho(\eta) = \sqrt{ \frac{\eps}{2 \pi} } e^{ - \eta^2 \eps / 2 }.
\ee
which implies that
\be
\int \rho( \eta) \eta^j d \eta= \left\{
                                  \begin{array}{ll}
                                   =1,              & j=0 \\
                                   =0,              & j=1  \\
                                   =1/\eps          & j=2  \\
                                   =O(\eps^{-j/2} ) & j \geq 3
                                  \end{array}
                           \right.          .
\ee

Let us denote  a probability density of finding the state with
given   $x$ and $p$ by $P(x,p)$. Updating the states gives a new
ensemble  with the probability distribution
\beq
P'(x',p') & = & \int dx \; dp \; P(x,p) P(x,p \rightarrow x',p')
\nonumber  \\
          & = & \int dx \; dp \; d \eta \; \rho(\eta) P(x,p)
\nonumber \\
   &  & \times \delta \left( p' - p - \eps \left[ -\frac{d V}{dx} -
          \frac{\alpha p}{m}
          +\left( \frac{2 \alpha}{\gamma} \right) ^{1/2} \eta (\tau)
          \right] \right)
\nonumber \\
   & & \times \delta(x' -x - \frac{\eps p'}{m} ) .
\eeq
A little algebra yields the result
\newcommand{\p}{\partial}
\beq
P'(x,p) & = & P(x,p) \nonumber \\
        & + & \eps \left[ \left( \frac{\p H}{\p x} \frac{\p P}{\p p}
              - \frac{\p H}{\p p} \frac{\p P}{\p x} \right)
              +\alpha \left( \frac{1}{\gamma}\frac{\p^2 P}{\p p^2}
              +\frac{p}{m} \frac{\p P}{\p p} +\frac{P}{m} \right)
               \right]
              +O(\eps^2) , \label{prob}
\eeq
whrere $H$ is the Hamiltonian corresponding to the original Newton's
equation of motion $ H= \frac{p^2}{2 m} +V(x)$.
Eq. (\ref{prob}) is a Fokker-Planck equation for the evolution of the
probability distribution $P(x,p)$. It is now readily verified to order
$\eps$ that  stationary distribution for such evolution is
the simple Boltzman weight
\be
P(x,p)= \exp \{ - \gamma H(p,x) \}
\ee
which can  be obtained through the very long evolution.

In the limit $m \rightarrow 0$ the Eq. (\ref{evol}) leads to the
Langevin equation in the rescaled time $ (\tau \rightarrow \alpha \tau)$
\beq
\frac{ d x}{d \tau}= - \frac{d V}{d x}
   + \left( \frac{2}{\gamma} \right) ^{1/2} \eta (\tau) .
\eeq
In the limit $\alpha \rightarrow 0$ one recovers the deterministic
limit called also microcanonical. Since in that case we get back
to the Newton's equation and the energy remains unchanged during
evolution the temperature should be measured by some sort of thermometer.
In our example it could be the average kinetic energy of the particle
$\frac{1}{2} kT= \la \frac{p^2}{2m} \ra$.
To change the temperature one should start from different initial
configuration.

Both approches: Langevin and microcanonical were used in numerical
simulations of fermions  \cite{langevin} \cite{micro}. In the standard
version Hybrid Monte Carlo (HMC) algorithm \cite{ken} makes use of
deterministic evolution equations.
For illustration purposes we consider the HMC algorithm for the
half-filled Hubbard Model. Introducing for every
$A_i$ the corresponding momentum $p_i$ the Hamiltonian
of the system can be written as
\be
H(A , p) = \frac{1}{2} \sum_i p_{i}^2 + \frac{1}{2}
             \sum_{i} A_{i}^{2} +\sum_{i,j} \phi_{i}^\star
             (M^\d M)_{i,j}^{-1} \phi_{j}  ,
\ee
where the indices $i$ and $j$ run over the whole space-time lattice.
Initially the gaussian momenta $p$ are chosen and fields $\phi$
are generated from gaussian vectors $r$ by $\phi = M^\d (A) r $.
Then the system evolves for $N_{mic}$ steps.
The evolution equations are:
\be
\stackrel{.}{A_{i}   } = \frac{\p H}{\p p_{i} } , \; \; \; 
\stackrel{.}{ p_{i} } = - \frac{\p H}{\p A_{i} } . \label{evoq}
\ee
The integration of Eqs. (\ref{evoq}) is performed according to the
leap-frog scheme
with a discrete step $\eps$, $N_{mic} \eps $ is the  trajectory length.
The conjugate gradient algorithm is required to compute the new vector
$(M^\d M)^{-1} \phi$ in each step of the evolution.

Due to the finite step errors the energy does not remain constant during
the evolution. Thus to ensure the exactness of the algorithm additional
global reject/accept step is needed. According to Metropolis scheme one
can fully restore the detailed balance by accepting the new
configuration $(p',A')$ with the probability
\be
P_{acc}= \min [ 1, \exp{ ( H(p,A) - H(p',A') ) } ]
\ee
The acceptance rate behaves  as
\be
P_{acc}= erfc(c N_{mic} \eps^3 \sqrt{V} ) \label{aa} ,
\ee
where $V$ is the total volume of the lattice and $c$ is a constant
factor. The errror function $erfc$ is defined as
\be
erfc(x) = \frac{2}{\sqrt{ \pi} } \int_{x}^{\infty} e^{-t^2} \; dt \; .
\ee
If the trajectory length is fixed the acceptance falls exponentially
with $\eps^4$. Thus to minimize the autocorrelations one should keep the
step size as large a possible still maintaining reasonable
acceptance.

Currently it is the most widely used algorithm in simulations
with dynamical quarks in QCD. In the optimal situation its
numerical complexity behaves  with the volume of the system as $V
V^{1/4} $, as can be readily
verified from Eq.\ (\ref{aa}) which is only slightly worse
than the linear dependence for bosonic algorithms. However these
algorithms are still rather complicated comparing with the simplicity
of pure bosonic algorithms and suffer from the strong autocorrelations
between generated configurations.
The proof of the detailed balance of the HMC algorithm requires exact
reversibility and it is unclear how finite numerical accuracy of
calculations can influence the reversibility of the algorithm and
further the physical results \cite{jan}. As can be seen later L\"uscher
method has no such difficulties. The resulting action is local
and systematic errors are under control.

\subsection{Algorithms for simulation of the Hubbard Model}
It is rather surprising that the algorithms described above
have not been widely used for simulations of the
Hubbard Model. Only recently extensive studies
of the attractive Hubbard Model has been performed with the help
of  HMC algorithm \cite{petersen}. Today's leading algorithm
\cite{white} is based rather on the exact numerical evaluation of the
determinants  appearing in the expression for the partition function.
Assuming that the two dimensional lattice with spatial volume $N=N_x N_y$
and $\nt$ time slices is considered, the partition function in Hirsch
formulation reads \cite{hirsch}
\be
Z= \sum_{s_{i, l}= \pm 1 } \det{ M^{+} (s)} \det{ M^{-} (s) }.
\ee
Here
\be
M^{\s}= I + B_{\nt}^{\s} B_{\nt -1 }^{\s}  \ldots B_{1}^{\s} \label{h}
\ee
and
\be
B_{l}^{\pm} = e^{\mp \Delta \tau \nu(l)} e^{-\Delta \tau \widehat{K} }
\ee
$I$ is the unit matrix, $\nu (l)_{i,j} = \delta_{i,j} s_{i, l}$ and
$\widehat{K}$ is the matrix representation of the kinetic part of
Hamiltonian. The $s$ is the Ising like Hubbard-Stratonovich field
which has two values $+1$ or $-1$. The
index $i$ labels the lattice sites and $l$ the time slices. The $M^{\s}$
matrices are $N \times N$ dimensional. A direct
evaluation of the determinant requires $O((N \nt)^3)$ operations.
Fortunately due to the very simple form of matrices there is a method of
updating the determinant with $O(N^3 \nt)$ operations needed to perform
full sweep through the lattice. The early implementations of this
algorithm suffered from numerical instabilities especially strong in
the limit of low temperatures and extrapolation to the ground state was
impossible. It turned out that
numerical instabilities occurred during multiplication of  the
badly conditioned matrices $B$ with almost linearly dependent columns.
A simple remedy was the periodic reorthogonalization of the fermionic
matrices and gave very good results.
At present simulations of the half--filled Hubbard Model can deliver results
with decent precision in the wide range of parameters.

An extremely difficult and not fully solved problem, referred to as the
sign problem, concerns the simulation of the fermionic system when the
determinant is not always positive.
This is the case of the Hubbard Model outside half-filled band.
A simple method for dealing with this difficulty
consists of attaching the unwanted sign of the probability
measure to the observable.
More precisely, if $P$ is not positively definite
then the expectation value of some observable $O$ can be written as
\be
\la O \ra_P = \frac{ \la O sign(P) \ra_{|P|} } { \la sign(P) \ra_{|P|} }
\ee
where $\la \ldots \ra_{ |P| } $ means the expectation value with the
respect to the absolute value of $P$.
Unfortunately the above equation, exact in principle, dramatically
increases statistical errors when the expectation value of sign
becomes  small. It is remarkable that the sign problem is especially
strong just below half-filling where there is a greatest chance of
finding the superconducting phase \cite{white}.

It seems that modification of the standard Quantum Monte Carlo
algorithms called Projector Quantum Monte Carlo will be
more useful for obtaining the ground state property of the Hubbard Model.
In this approach, proposed in Ref. \cite{kon} for bosonic systems and
applied further to the Hubbard Model \cite{car}, the projected partition
function is introduced
\be
Q= \langle \psi_T | \exp{ [- \beta {\cal H}] } | \psi_T \rangle
\ee
$|\psi_T \rangle $ is a trial wave function nonorthogonal to the
ground state (usually the Slater determinant). The Hamiltonian is used
to systematically project the trial wave function onto the ground state.
Thus in the limit of large imaginary time $\beta$ the
properties of the ground state can be obtained.
The quantity $Q$ can be evaluated by the Monte Carlo methods
after rewriting it as a path integral.

The very interesting point is the possibility of circumventing the sign
problem in this approach. It can be achieved by the appropriate choice
of the trial function.  It can be also handled by applying another
statistical weight which
gives the identical distribution in the low temperature limit
with the reasonable expectation value of sign.
The construction of better probabilistic weights is rather problem
dependent, some simple examples are given in Ref. \cite{car}.
Recently similar philosophy has been applied to the noninteracting
electron system  with chemical potential \cite{sign}.
The desired distribution is constructed iteratively with the help
of the cluster algorithm. It was shown that for this simple model
the sign problem can be completely eliminated. Of course it would be
of great interest to extend this method to more complicated models.

Efficiency of the Projector Quantum Monte Carlo algorithm
depends on the type of the ground state for noninteracting
system (U=0). Usually if the ground state is unique
the convergence is easily achieved. Conversely in the case of the
degenerate ground state  the algorithm suffers from the
negative sign problem and poor statistics.

According to Eq. (\ref{disp}) the allowed states in the momentum space
can be divided into groups (shells) with a given energy. Degeneracy in the
ground state appear as a result of the existence of not completely
occupied shell.  More precisely, if $m$ electrons with the specified
polarization occupy the shell which can contain maximally up to $n$ such
electrons it leads to a degeneracy
\be
N_d = \left( \begin{array}{c}      n \\ m \end{array} \right) .
\ee
Although there is no analytical expression which gives the
degeneracy as a function of lattice size and total number of electrons
it is easy to compute this degeneracy numerically dividing the states
in the momentum space into shells.
As an example we consider a system with almost half-filled band, 7 electrons
up and 7 electrons down on the $4 \times 4 $ lattice. The ground state
has a degeneracy of $N_d=29$, as far as the sub-space of total
momentum $Q=0$ is considered. Therefore we may expect
rather poor efficiency of PQMC algorithm in this physically important case.

Quite different numerical approach is based on the exact diagonalization
of many particle Hamiltonian.  The main problem is the enormous
dimension of the Fock space $D_{Fock}=4^N$.
Fortunately symmetries of the problem often significantly reduce this number.
Although the dimension of matrices precludes their
direct diagonalization it is relatively easy to obtain
eigenvectors for few largest and smallest eigenvalues using Lanczos
method \cite{lanczos}. They contain all necessary information needed to
compute physical observables. Additional very nice feature
of this method is the ability to obtain the interesting
quantities directly in the Minkowski time (real frequency properties).
In Quantum Monte Carlo calculations are carried out
in imaginary time and the statistical errors inherent to the
Monte Carlo make the analytical continuation difficult.

\subsection{Local bosonic algorithm}

Recent  proposal of L\"uscher has attracted a lot of interest \cite{lu}.
Originally being introduced as an alternative to the Hybrid Monte Carlo
algorithms in QCD, soon has been applied in QMC simulations of the Hubbard
Model \cite{saw}. The basic idea of L\"uscher consists in approximating of
the inverse of the fermionic matrix with a polynomial $P_{n}$ of even
degree. Its particular form being proposed is  built from Chebyshev
polynomials and  gives rapid uniform convergence to the function $1/x$
in the interval $( \eps, 1 )$, with $\eps$ being a small positive number
\be
\lim_{n \rightarrow \infty} P_{n}(x)=1/x. \label{polz}
\ee
The relative error of the approximation defined as $R_n(x)=x P_n (x)-1$
is exponentially bounded
\beq
   |R_n(x)| < 2  \left( \frac{ 1- \sqrt{ \epsilon}}{1+ \sqrt{ \epsilon}}
              \right)^{n+1}. \label{bound}
\eeq
Approximation region should cover the whole spectrum of the fermionic
matrix or at least its significant part. Therefore it is impossible to apply
directly the original L\"uscher polynomial $P_n$ to the
nonhermitian matrix $M$ with eigenvalues distributed on the complex
plane. Instead, we wrote the polynomial
approximation for the matrix $Q^\d Q = M^\d M / \lambda_{max}$,
where $\lambda_{max}$ is the largest eigenvalue of $  M^\d M $.
Hermitian and positive definite matrix $Q^\d Q$ has properly distributed
eigenvalues in  the interval $(0,1)$.
\beq
\frac{1}{ Q^\d Q}= P_{2n} (Q^\d Q)= \prod_{k=1}^{2n}
( Q^\d Q -z_{k} ) \nonumber \\
= \prod_{k=1}^{n} ( Q^\d Q - \alpha_k -i \beta_k)
( Q^\d Q - \alpha_k +i \beta_k)
\eeq
since the roots $z_k= \alpha_k +i \beta_k$ of real polynomial
$P_{2n}$ come in complex conjugate pairs. Their values are analytically
known with the accuracy of $O(1/n)$. More precise values can be simply
obtained by applying few Newton-Raphson iterations. After introducing
auxiliary fields $\phi_k$ the local bosonic representation  for
the partition function of the Hubbard Model reads
\be
  Z  \simeq  \int   [dA d\phi] e^{-S} , \label{zet}
\ee
where
\be
S = \sum_i A_{i}^{2} /2    + \sum_{i,j}  \sum_{k=1}^{n}
\phi^{\star}_{k,i} \left[ ( Q^{\dagger} Q -\alpha_k)^2+\beta_{k}^{2}
\right]_{i j} \phi_{k,j}     \label{akcja}
\ee
with $M$ given by Eq. (\ref{matrix}).Here the indices $i$ and $j$ run
over the whole space-time lattice with volume $V=N_{x}^{2} \times \nt$.
The action
can be rewritten also directly in terms of the real  and the imaginary
part of $\phi_{k,i}=\phi_{k,i}^{(0)} + i\phi_{k,i}^{(1)}$
\be
S = \sum_i A_{i}^{2} /2    + \sum_{i,j}  \sum_{r=0,1} \sum_{k=1}^{n}
\phi^{(r)}_{k,i} \left[ ( Q^{\dagger} Q -\alpha_k)^2+\beta_{k}^{2}
\right]_{i j} \phi_{k,j}^{(r)}.
\ee

Some remarks are in order. The resulting action $S$
contains only local interactions. Hence we can study the fermionic system
using bosonic Monte Carlo techniques. However some complications are
introduced by the  presence of next-to-nearest neighbors interactions.
The condition number of matrix $Q^{\dagger} Q$ is square of
condition number for $M$ and we may
generally expect  the poorer behaviour of numerical procedures. Moreover it
complicates the implementation of this algorithm on vector and parallel
computers. The analogical action for QCD is simpler because one make use
of additional symmetry of the Dirac operator.

To define the normalized matrix $Q$ we used the upper bound for the
$\lambda_{max}$ which can be easily derived
\be
\lambda_{max} < 2[ \lambda_{max} (A^\d A) +\max(D)^2 ] \label{error}
\ee
where $D$ is diagonal part of $M = A+D$. Large magnitudes of the $A$ fields
are  naturally cut by the gaussian part of the action and we may safely
assume maximal value for the diagonal elements; e.\ g.\ the $A(x) <4$ would
correspond to four standard deviations assuming that the gaussian part
of the action dominates.

\subsection{ Exact algorithm}
\newcommand {\dm}{ \Delta M^{-1} }
Below we discribe algorithm based on direct computation of
fermionic determinant. We use this algorithm as a kind of reference
point which, first allows us by comparison to control the accuracy
of the L\"uscher algorithm, and second gives us a chance to compare
an efficiency.

The idea of exact algorithm  is  based on the updating
scheme allowing to compute changes in $M^{-1}$ matrix while making
trial changes in matrix $M$.  In order to update the field $A_k$ located
on the site $k$ one must calculate the ratio of the
fermion determinants. Since the only $A$ dependent elements of matrix $M$
lie on the diagonal we consider the following change in the matrix $M$
\be
M'= M + \Delta ,
\ee
where $\Delta$ is matrix with one nonzero element, say k-th on the
diagonal,
$\Delta_{i j}= \delta_{i k}  \delta_{k j} d$. Then
\be
\frac{ \det {M'} } { \det {M} }= \det{ (I + M^{-1} \Delta) }
= 1 + M^{-1}_{k k} d
\ee
is determined completely by elements of matrix $M^{-1}$. Once the trial
change has been accepted the updated $M^{-1}$ can be evaluated
from the Shermann-Morrison formula \cite{nr}
\be
 M^{' -1}= M^{-1} - \frac{M^{-1} \dm}{1 +d M^{-1}_{k k} }.
\ee

This process is  economical from the computational point of view since one
update of $M^{-1}$ requires $O(V^2)$ operations  comparing with $O(V^3)$
operations needed  to evaluate the determinant with the brute force method.
Today's leading QMC simulations of the Hubbard Model are essentially
based on it \cite{white}. As has been previously reported numerical
instabilities often  appear
in such calculations especially at low temperatures. However our particular
formulation of path integrals does not require  multiplication of badly
conditioned matrices and we believe that it is free of this difficulty.
Indeed we performed thousands sweeps at $\beta$ as large as 8 without
accumulating numerical errors. However the additional cost comes from
working with larger matrices.

\section{Results}

In this section we present implementation details of the L\"uscher algorithm.
We compare the accuracy and efficiency of the local bosonic algorithm with
the exact algorithm described in the previous section.
We use these algorithms to study basic features of the Hubbard Model.
In our studies we looked mainly at the magnetic properties
of the Hubbard Model and one-particle Green's functions (shape of the
Fermi surface and effective  hopping). Due to the sign problem our
discussion is restricted to the half-filled band.

\subsection{Implementation details and the errors of
             the polynomial approximation}

The goal of the MC simulation is to generate $A$ and $\phi$ fields
with the disribution (\ref{zet}). To achieve this goal we alternately
update all components of $A$ and all components of $\phi$. We first
consider the update of $\phi$ fields. Because of the gaussian form of
$\phi$ dependent part of the action in Eq. (\ref{zet}) we applied the
heat bath algorithm \cite{sokal} to update this sector. This is
implemented as follows. For  every $\phi_{k,i}^{(r)}$ component we
compute the conditional parameters of its gaussian distribution while
keeping other components fixed. Then the value  obtained from gaussian
random number  generator is assigned to the $\phi_{k,i}^{(r)}$. In a
full sweep through the lattice these operations are performed for each
$i$, $r$ and $k$.

The effective action depends on $A$ in more complicated way and and
the heat--bath algorithm is not longer useful. Instead we use a more
general Metropolis algorithm \cite{metro}. The trial part of a single
update relies on proposing a change
\be
A_i = A_i + a ( x -0.5).
\ee
The Metropolis  step accepts the proposed change with the probability
\linebreak $\min (e^{-\Delta S},1)$, where $\Delta S$ is corresponding
change of the action. This guarantees the detailed balance condition.
$x$ is a random number uniformly  distributed in the
interval $(0,1)$. The parameter $a$  introduces the scale of changes and it
is very important to tune it to get minimal autocorrelation times.  Usually
it is choosen to keep the acceptance ratio  near $50 \% $ . It is also very
important to maintain the balance between updates of $A$ and $\phi$ fields.
Earlier studies of QCD implementation suggest that one should perform
updates of the $A$ field much more frequently than the $\phi$ fields to 
minimize
autocorrelation times \cite{jeg}. In our simulations ten Metropolis
updates of $A$ field follow one heat-bath generation of $\phi$ fields.

Since the CPU cost is expected to be at least proportional to the number of
fields $n$ it is important to adjust this number carefully.
One sees from the Eq. (\ref{bound}) that to decrease $\eps$ one has
to increase $n$ to keep the error of approximation constant. In the case
of the approximation of the determinant of the matrix $Q^\d Q$ one can
easily guess that the most economic choice of $\eps$ should be
comparable to the smallest eigenvalue
of the matrix $Q^\d Q$. To make this discussion more precise we introduce
as a measure of the error of the polynomial approximation the quantity
\cite{for2}
\be
\delta=| y^{1/V} -1|
\ee
with
\be
y= \det{ ( Q^\dagger Q P_{n}(Q^\dagger Q) )  }
\ee
The power $1/V$ properly normalizes the quantity $y$ per one degree of
freedom.
\fc shows $\delta$ as a function of $\eps$ for few numbers of fields
measured on  one typical configuration and confirms simple
expectations ($\lambda_{min} =0.0032$ ). With this hint we estimated
the number of fields $n$ to be about 50-100. This guaranteed that the
relative error $|R_{2n}|$ was smaller than $10^{-4}$ in the whole
interval $(\eps, 1)$. If one accepts bigger relative errors,
for example $10^{-2}$ these numbers can be
reduced but in this case we found discrepancies between exact algorithm
and L\"uscher implementation (Table.1).
\begin{table}
   \begin{center}
 \begin{tabular}{|c|c|c|c|}
 \hline
   $ |R_{2n}(x)|  $          & $10^{-2}$  & $10^{-3}$ & $10^{-4}$ \\
 \hline
  $\eps =0.003$ & 26         & 34        &  45       \\
  $\eps =0.001$ & 41         & 60        &  78       \\
  $\eps =0.0005$ & 59         & 84       &  110       \\
 \hline
  \end{tabular}
 \caption{ {\small The number of fields needed to achieve different
levels of relative errors $|R_{2n}(x) |$ on the interval $(\eps ,1)$ } }
  \end{center}
\end{table}

\subsection{Dynamical properties of the L\"uscher algorithm and
                      its modifications }

In order to establish the computational effort required to
generate one uncorrelated (independent) statistical event we studied
the integrated autocorrelation times for different observables.
Let $\{ O_t \} (t=1, \ldots ,N_s) $ be an ordered sequence  of data
produced by our algorithm with the mean value $\mu$. We define the
unnormalized autocorrelation function
\be
C(t)= \la O_s O_{s+t} \ra -\mu^2,
\ee
normalized autocorrelation function
\be
\rho (t)=C(t)/C(0),
\ee
and integrated autocorrelation time
\be
\tau_{int} = \frac{1}{2} \sum_{t= -\infty}^{\infty} \rho (t).
\ee
The integrated autocorrelation time controls the statistical
error of the sample mean
\be
\hat{ \mu } = \frac{1}{N_s} \sum_{t=1}^{N_s} O_t .
\ee
Indeed its variance can be rewritten for sufficiently  large
samples as
\be
var( \hat{ \mu} ) = \la ( \hat{ \mu} -\mu )^2 \ra = \frac{1}{N_{s}^{2} }
\sum_{r ,s =1}^{N_s} C( r-s) \simeq \frac{1}{N_s} \
\sum_{t = -\infty }^{\infty}
C(t) = \frac{1}{N_s} \; 2 \tau_{int} C(0).
\ee
Therefore the $2 \tau_{int}$ is the number of iterations
needed to produce one uncorrelated estimate for $O$.

Obviously in real situation the estimation of $\tau_{int}$
have to be done from a finite data set. However summation
over  all available data is misleading since the tail of $\rho$
contains a lot of noise but little signal.
Therefore in practice  we sum the autocorrelations only
up to a certain distance $T_{cut}$
\be
\tau_{int}(T_{ cut} ) = \frac{1}{2} + \sum_{t=1}^{T_{cut} } \rho(t) .
\ee
There are two limitations. If $T_{cut}$ is too large our estimator
is very noisy since its variance is
\be
var(\tau (T_{cut})) = \frac{2 (2 T_{cut} +1)}{N_s} \tau_{int}^{2}.
\ee
On the other hand two small value of $T_{cut}$ introduces
a significant bias. The optimal value of $T_{cut}$ can be easily estimated
from sufficiently long runs by the ``window'' procedure \cite{aut}.
The recipe is  simple: choose $T_{ cut} $ to be the smallest integer
such that $T_{cut} \geq c \tau_{int} (T_{cut}) $.
Assuming pure exponential behaviour
of $\rho(t)=e^{- t / \tau_{int} }$ it is sufficient to take $c=4$.
Then the bias of our estimator is of order of $e^{-4} = 2 \% $ .

We have calculated the autocorrelation times for two observables:  the
average density of electrons with spin up $\la n_{i \uparrow} \ra $, and
that of pairs of electrons with the opposite spins on the same site $\la
n_{i \uparrow} n_{i \downarrow} \ra $ . The simple updating scheme
described above gives large autocorrelation times already for $K=1$ and
$\beta=1$.  It is important to learn whether the long autocorrelations
appear while generating $\phi$ fields or during update of the $A$ field. We
have introduced some modifications to the algorithm to address this
question.

\begin{table*}[hbt]
\setlength{\tabcolsep}{0.4pc}

\newlength{\digitwidth} \settowidth{\digitwidth}{\rm 0}
\catcode`?=\active \def?{\kern\digitwidth}
\caption{}
\label{tab:effluents}
\begin{tabular*}{\textwidth}{@{}l@{\extracolsep{\fill}}rrrr}
\hline
                 & \multicolumn{1}{l}{$\beta=1$, $U=1$ }
                 & \multicolumn{1}{l}{$\beta=1$, $U=1$ }
                 & \multicolumn{1}{l}{$\beta=1$, $U=2$ }
                 & \multicolumn{1}{l}{$\beta=1$, $U=2$ } \\
                 & \multicolumn{1}{c}{lattice $5^3$  }
                 & \multicolumn{1}{c}{lattice $6^2 8$}
                 & \multicolumn{1}{c}{lattice $6^2 8$  }
                 & \multicolumn{1}{c}{lattice $6^2 14$  }  \\
\hline
Exact determinant & 0.460(2) & $0.473(2)$ & $  0.462(4)$ & 0.468(5)   \\
                   &  0.2197(2) & $ 0.2203(4)$ & $0.195(1)$ & 0.193(1) \\
                  & $\tau_1=3 $ & $\tau_1=3 $ & $\tau_1=3 $& $\tau_1=2 $ \\
                  & $\tau_2=1.5 $ & $\tau_2=1.5 $ & $\tau_2=1.5$ & $\tau_2=1$ \\
\hline
Simple program     & 0.461(5)  &  $0.471(6)  $ & --- & ---  \\
                    &  0.2180(8)  & $ 0.221(1) $ &  &    \\
                    & $\tau_1=660$ & $\tau_1=870 $ & &    \\
                    & $\tau_2=320$ & $\tau_2=300 $ & &     \\
\hline
Simple program     & --- &  $0.470(4)$      &  ---      &  ---  \\
with MG (W-cycle)  & & $0.2194(6)$    & &  \\
                   & & $\tau_1=160$  &  & \\
                   & & $\tau_2=80$  &  & \\
\hline
Global generation & $0.449(7)$&--- &--- &--- \\
of gaussian fields  & $0.221(1)$ & & &  \\
                      &  $\tau_1=180 $ & & &  \\
                      &  $\tau_2=90 $ & & &  \\
\hline
Simple program    & --- & $0.470(5)$ &  $0.463(7)$  & $0.475(7)$  \\
with preconditioning & & $0.220(1)$ & $0.195(1)$   & $0.186(3)$   \\
        & & $\tau_1 =100$ & $\tau_1 \simeq 200$  & $\tau_1 \simeq 600 $ \\
        & & $ \tau_2=60 $ & $\tau_2 \simeq 200$ & $\tau_2 \simeq 600 $ \\
\hline
\multicolumn{5}{@{}p{148mm}}{ \mbox{} } \\
\multicolumn{5}{@{}p{148mm}}{{ \small The results for density of electrons
$\la n_{i \uparrow} \ra $ (first line) and density of pairs $\la n_{i
\uparrow} n_{i \downarrow} \ra $ (second line).
The autocorrelation times are given for both quantities. Autocorrelation
times for exact numerical evaluation of the  determinant are of the
order of unity.}   }
\end{tabular*}
\end{table*}

One could think that these correlations are caused mainly by the critical
slowing down introduced by $\phi$ fields, especially those with small
$\beta_k$. This would suggest that one would get a better algorithm
efficiency by improving updating of the gaussian sector. There are many
efficient techniques used in simulations of the gaussian fields. Among
them the Multigrid algorithm eliminates completely critical slowing down in
the pure gaussian model and therefore we decided to use it to update
$\phi$ fields in our problem \cite{sokal} \cite{ed}. In this approach
one considers the sequence of coarse-grid problems  which approximate
original problem on different  length scales and the  local updates of
the heat-bath algorithm are supplemented by collective updates.
We define the set of blocks $B_i$. For cubic lattice the blocks are taken
successively to be single sites ($B_0$ ), cubes of side 2 ($B_1$ ), cubes
of side 4 ($B_2$) and so on. Obviously if the lattice size is not a power
of $2$ we can choose also other small blocking factor. With every block
we associate the conditional probability distribution, which for the
$\phi_{k}^{(r)}$ field reads
\be
P_{B} (t) \propto \exp{ \left[ -\sum_{i,j}  (\phi_{k,i}^{ (r) } + t \,
(\chi_B)_i )
\left[ ( Q^{\dagger} Q -\alpha_k)^2 +\beta_{k}^{2}
\right]_{i,j} (\phi_{k,j}^{ (r) } + t \, (\chi_B)_j ) \right]  }.
\label{pcond}
\ee
$\chi_B$ denotes the vector whose components are 1 for sites belonging to
the block $B$ and 0 elsewhere. The $P_{B}$ depends only on one real
number $t$.

This general set up can be directly used in our problem as follows. We
begin with elementary sites for which we perform
the full heat-bath sweep through the lattice sequentially updating
$\phi_{k,i}^{(r)}$ for all $i$, $k$ and $r$. Then we organize the lattice
into blocks $B_1$. We change the $\phi_{k,i}^{(r)}$ at all sites of
given block by the same $t$ according
to distribution (\ref{pcond}). After having done the sweep over blocks
$B_{1}$, for all $k$ and $r$ we repeat the same for blocks $B_2$  and so on.
This updating scheme is called V-cycle. The sweeps can be also organized
in such fashion that there are $\gamma^{l}$ updates of blocks $B_l$.
The control parameter $\gamma$ defines different classes of MG algorithm.
The most common choices are $\gamma=1$ (V-cycle) and $\gamma=2$ (W-cycle)

Indeed the MG generation of $\phi$ fields reduces the autocorrelation times
substantially (third row of Table 2). However it introduces additional
computational cost. In fact the CPU time required to get one independent
sample is even bigger than for a simple heat-bath method.  The maximal
decorrelation of auxiliary fields is achieved by the  independent
generation of the eigenmodes. This requires the solution of set of linear
equations for each field $\phi$ and the results are comparable to the MGMC
case with W-cycle (4-th row of Table 2).

Neither simple version nor MG refinement is capable to reproduce the exact
determinant results for $U=2$. The system did not thermalize even after 20
times more thermalization steps than required for $U=1$. This fact can be
simply understood. When $U$ increases the condition number of $M^\dagger M$
becomes bigger and  larger number of  $\phi$ fields in polynomial
approximation is needed. Additionally they are stronger coupled to the $A$
field because of the presence of factor $\exp{\sqrt{U \beta /\nt}}$. The
constraints imposed by large number of $\phi$  fields on one $A$
field becomes more restrictive and the mobility of algorithm rapidly
decreases. Performing updates of $A$ field more frequently is only a
partial solution.

Better results gives the preconditioning of fermionic matrix. We define the
preconditioning procedure as follows. Let $D$ be a diagonal part of
$M=A+D$. Then the  matrix $M^\dagger D^{-1} M$ is better conditioned
than $M^\d M$ as can be  seen from \fc . The effect is clearly visible
in the last row of Table 2. As a consequence wider range of couplings
become available. However the reliable runs at $U$ greater than 2 are
not feasible.

The program based on exact evaluation of determinant has no such restrictions
and works equally good at $U=1$ like for $U=4$. Required CPU time
needed for both algorithms varies with lattice sizes. On small
lattices algorithm based on exact evaluation of determinant
is substantially faster and it took 20 minutes on HP735/125 workstation
to obtain the results on lattice $6^2 8$ while comparing with $10$ hours
for L\"usher implementation. Of course for larger lattices the
computational effort grows much more faster for exact algorithm.
In the region of weak coupling and high temperatures we were able to perform
simulations on lattices $16^2 \times 8$ with the help of the L\"uscher
algorithm.

The polynomial approximation can be extended simply to
the complex plane \cite{monty} \cite{for2} i.\ e.\ for nonhermitian
matrices. Thus one tries to approximate directly
the inverse of matrix $M$. This would reduce the condition numbers
for matrices entering the problem and would result in the simpler final
action. This modification of original L\"uscher idea gave very promising
results \cite{for2} \cite{galli}.  However in contrast to the QCD case the 
matrix $M$
for the Hubbard Model has eigenvalues with the positive as well
with negative real parts as can for example be seen on  \fc.
Because one cannot extend the
domain of the applicability of Eq. (\ref{polz}) beyond the singular
point $(0,0)$, it is unfortunately
impossible to adapt this modification to the Hubbard Model.

On the other hand we have found that one change the positions of
eigenvalues by introducing the chemical potential. In fact we managed
to shift all eigenvalues of the matrix $M$ in simulations
of the  {\em attractive } Hubbard Model to the right half of the complex
plane. However it occurred at large $\mu$ which corresponds to the
physically uninteresting filling ($\la n_\uparrow +n_\downarrow \ra >0.9$).

Other  improvements of the L\"uscher algorithm have been proposed.
Global Metropolis step \cite{for1} makes the algorithm exact and
heat-bath generation of $\phi$ fields with overrelaxation slightly reduce
autocorrelation times (for review see \cite{adler}). We believe  however
that it would not change the situation qualitatively.

\subsection{Magnetic properties of the Hubbard Model}
In section 2 we argued for the existence of the antiferromagnetic order
in the ground state of the Hubbard Model. These arguments were based
on the mean-field approximation and should be independently confirmed in the
numerical simulations. In fact, the mean field theory is known to
overestimate the magnetic ordering and to underestimate quantum
fluctuations.  The situation is rather complicated since
the Hubbard Model has continuous (spin - SU(2)) symmetry. In that case,
in $2d$, the Mermin-Wagner \cite{mermin} \cite{aza} theorem
prevents the existence of the long--range magnetic order at finite
temperature. The correlation length can become really infinite only at
zero temperature when  thermal fluctuations vanish.

Therefore to address the problem of magnetic properties of the Hubbard Model
it is necessary to perform simulations at large $\beta$.
This in turn requires sufficient $\nt$ in the discretized form of
the path integral  formulation to reduce the systematic errors of order
of $(\beta/\nt)^2$. Moreover in our particular representation the number
of electrons with spin down and with spin up differ. It is special
inconvenience in measurements some correlation functions leading
to the errors which are extensive function of the lattice size.

To estimate the temperature we can reach, in the physical units we
assume that $8K=1eV$. Then it follows from Eq. (\ref{disp}) that
$\beta=5$ and $K=1$, the most typical values for our runs correspond
to temperature $T= \frac{1}{40} eV =290$ Kelvin degrees.   
It would be difficult to reach much lower temperatures due to 
computer restrictions and finite $\nt$ effects.

We begin discussion with a local quantity which can be simply measured
on small lattices with good accuracy. The average number of pairs
with opposite spins $\la n_{\uparrow} n_{\downarrow} \ra $ on the
same site shows how repulsive interaction forbids double occupancy of a
site. Clearly   increasing $U$ leads to reduction in number of pairs.
\fc compares  results from QMC simulations with the mean field
prediction. The  derivation of the latter is presented in the Appendix
A. Here we note a good qualitative agreement with MC results. For this
quantity a $4 \times 4 $ lattice gives rather good
approximation to the bulk result and we did not notice meaningful
finite size effects as can be seen from Table.3 .
\begin{table}
\vspace{1cm}
\begin{tabular}{| c || c | c | c | c| c ||}
\hline
$ \la  n_{i \uparrow} n_{i \downarrow} \ra $ & N=16 & N=36 & N=64 & N=100
& N=144 \\
\hline
$ U=1  \; \beta =1  \; \nt=8 $ & 0.2195(4) & 0.2204(4) & 0.2202(5)
& 0.2202(3) & 0.2198(3) \\
\hline
$ U=4 \;  \beta=5 \;  \nt=30$ & 0.134(2) & 0.138(2) & & &  \\
\hline
\end{tabular}

\caption{ {\small  The density of pairs on the same site as a
function of the spatial volume of the lattice $N=N_{x}^{2}$. 
For this quantity the lattice $4^2$ already gives good approximation to
 the bulk result }
}
\end{table}

To decide weather the ground state has the long range antiferromagnetic
order one can attempt to measure  the equal time correlations
between magnetization on different sites
\be
C(l_x, l_y)= \la (n_{i \uparrow} - n_{i \downarrow})
(n_{i+l \uparrow} - n_{i+l \downarrow}) \ra.
\ee
This quantity obtained form simulations on $4 \times 4 $ lattice at
$\beta=5$ and $U=4$ is shown on \fc. The visible zig-zag is an
indication of the onset of the antiferromagnetism. We found that this
observable is very sensitive to finite $\nt$ effects. We needed 60 time
slices to obtain clear signal (the analogical zig-zags for $\nt=30$ and
$\nt=40$ do not look so nice). This is not surprising since the
inequality between $\la n_{i \uparrow} \ra $ and $\la n_{i \downarrow}
\ra$ is equivalent to putting the system into a magnetic field which
destroys the antiferromagnetic correlations. Within the spin wave theory the
correlation between two most distant points on the lattice is simply
related to the antiferromagnetic order parameter $m$ \cite{huse}
\be
C(N_x /2, N_x /2) = \frac{m^2}{3} + O(1/N_x)
\ee
with the finite size corrections of order $O(1/N_x)$ on the lattice with
spatial volume $N=N_{x}^{2}$. For the parameters of run the mean
field
result $m^2 /3 =0.15 $  should be compared with the value $0.10 \pm
0.04$ read from  \addtocounter{x}{-1} \fc. Thus even for relatively high
temperature of our simulations we obtained a good agreement. We did not
estimate the finite size effects because it requires the simulations on
bigger lattices.

Fourier transformation of $C(l)$ gives the magnetic structure factor
\be
S(q)= \sum_{l_x, l_y} \exp{ (i q l) } C(l)
\ee
which is especially well suited for detecting  excitations with wave vector
$q$. In the presence of antiferromagnetic long-range order we expect
divergence of the $S(\pi, \pi)$ with the lattice size according to
\be
S(\pi, \pi)= N \frac{m^2}{3} +O(1/N_x)   \label{spin}.
\ee
So far no one has shown in direct simulations the existence of
antiferromagnetic phase at $U$ below 2 \cite{anti}. Since we can reach
quite large lattices at small U using the L\"uscher method we made
such an attempt. \fc gives the  $S(k_x, k_x)$ as a function of momentum
on lattices with spatial sizes ranging from $4 \times 4$ to $16 \times
16 $ for $\beta=1$ and $U=1$.  Within the statistical errors there is no
dependence of the magnetic structure factor on lattice volume at this 
temperature.  An exception is $S(0,0)$ which rises with $N$ almost  
linearly. It is a simple artefact of
finite $N_t$  and does not mean  that system has ferromagnetic long
range order. Indeed the $S(0,0)$ is
defined as an extensive quantity and naively we can expect the 
contribution to the $S(0,0)$ of order of $N (\Delta n)^2$ coming
 from the difference
$\Delta n$  between $\la n_\uparrow \ra$ and
$ \la n_\downarrow \ra$. This crude estimation is tested in \fc
where the $S(0,0) $ is shown as a function of the spatial  volume of the
lattice for $\nt=6$, $\nt=8$ and $\nt=12$. For each $\nt$ we found
the parameters of linear fit
\be
S(0,0) = b N +a    \label{s00}
\ee
The results are listed in Table 4. The values of $b$ should be compared
with the mean difference $(\Delta n)^2$. The remaining contribution
$a$, which does not depend on $\nt$ may be regarded as a true $S(0,0)$.
The nonozero value of $S(0,0)$ is an effect of the short-range correlations.

\begin{table}
\vspace{0.5cm}
\begin{center}
\begin{tabular}{| c | c | c | c | }
\hline
 & $\nt =6 $ & $\nt=8 $ & $\nt=12$ \\
\hline
 a = & 0.33(3) & 0.34(2) & 0.34(3) \\
\hline
 b= & 0.044(4)  & 0.030(2) & 0.013(2) \\
\hline
$(\Delta n)^2$ &0.041  & 0.025 & 0.010  \\
\hline
\end{tabular}
\end{center}

\caption{ {\small  Parameters  $a$ and $b$ obtained from
simulations on lattices with $\nt =6,8,12$. The mean difference
$(\Delta n )^2$ is also given. } }

\end{table}

The simulations described above were performed at rather high temperature
(corresponding $\beta =1$). We did not found the antiferromagnetism
because of the disordering effect of thermal fluctuations. We performed some
simulations with the help of the exact algorithm at $U=4$ and the
$\beta=5$ where the scaling predicted by the Eq. (\ref{spin}) has been
observed \cite{scalapino}. Due to strong finite $\nt$ effects to obtain
valid result one has to perform extrapolation.  The finite $\nt$ errors
introduced by our lattice discretization are of the order of
$(\frac{\beta}{\nt})^2$. In \fc we present $S(\pi, \pi)$ as a function
of parameter  $(\frac{\beta}{\nt})^2$. In the limit $1/\nt \rightarrow
0$  for the lattice
$4\times 4$ we obtained $2.6 \pm 0.2$ and for the lattice $6 \times 6$ $3.2
\pm 1.0 $.   They compare satisfactorily  with MC
simulations of Ref.\ \cite{scalapino} (see Table. 5).
\begin{table}
  \begin{center}
    \begin{tabular}{| c | c | } \hline
    N=16 & N=36  \\\hline
     $2.6 \pm 0.2$ &  $3.2 \pm 1.0 $ \\\hline
     2.8 &  4.2 \\\hline
\end{tabular}
\end{center}

\caption{ {\small Comparison of results for $S(\pi, \pi)$ obtained
from our studies (first row) with those from Ref. [12] (second row). } }

\end{table} 

\subsection{Fermi surface}
It is also very interesting to study information contained in one
particle finite temperature Green's function defined in the momentum
space and imaginary time as
\be
G(k, \tau)= - \la  T a_{k \s} (\tau) a_{k \s}^{\d} (0)  \ra ,
\ee
where $a(\tau)$ is the annihilation operator in the Heisenberg picture
and $T$ time ordering operator. The limit of $G(k, \tau \rightarrow 0^-
)$ has simple physical content. It gives distribution of electrons in
the momentum space.  To see the shape of the Fermi surface one
has to perform simulations on sufficiently large lattices.
The largest lattice accessible to our simulations was $8^2 \times 20$.
For the parameters of run $U=4$ and $\beta=5$ the number of time slices
might appear too small. However, for this quantity we do not expect the
presence of the mechanism described above which leads to the errors
which are extensive function of the spatial lattice volume.

Finally, one can also compare our results with the mean-field
prediction. It follows from Eq. (\ref{ocup}) in Appendix A that 
\be
n(k)= n_{k \downarrow} + n_{k \uparrow} =
1 - \frac{\eps_k}{\sqrt{ \eps_{k}^{2} + \Delta^2} }
\ee
The energy gap $\Delta= 1.38$ is almost the same for the lattice
$6 \times 6$ and for the lattice $8 \times 8$.  Consequently we can
together compare results from lattices $6 \times 6$ and $8 \times 8$  
with the mean field formula (solid line in \fc ). Even for
our relatively  large $\beta/\nt =0.25$ the results we obtained are
in a good agreement with the theoretical prediction. However the
Fermi surface of the interacting system is more smooth than in the
noninteracting case (dotted line). Obviously this effect is an
increasing function of $U$.

The lack of sharp Fermi surface suggests vanishing $Z$ and behaviour
different from the Fermi liquid. Of course it would be very interesting
to measure the $n(k)$ at non zero chemical potential $\mu$ and find
the critical value $\mu_{crit}$ when the sharp Fermi surface appears.
Unfortunately the QMC simulations are not conclusive at small $\mu$
due to sign problem.  Currently it is only confirmed that
$\mu_{crit}$ is small.

\subsection{Effective hopping}
As pointed out in various approximate schemes the interaction leads
to the reduction in the effective hopping. In our Monte Carlo
simulations we measured the ratio
\be
\frac{K_{eff}}{K_0 }= \frac{  { \la a_{i \s}^{\d} a_{j \s} \ra }_U   }
                        {  { \la a_{i \s}^{\d} a_{j \s} \ra }_0   },
\ee
which shows how interaction reduces the effective hopping matrix elements
between nearest neighbor sites $i, j$. This quantity can be measured
directly or can be expressed in terms of Green's function.
Indeed $K_{eff}$ can be written as an expectation value
over the whole lattice
\be
K_{eff} = \frac{1}{4 N} \la \; \sum_{x, \nu} a_{x}^{\dagger} 
a_{x + \nu} + h.c \;  \ra,
\ee
where $x$ denotes the sites on the lattice, $\nu$ is the unit versor 
and $N$ is spatial volume of the lattice.
For simplicity we drop the spin index. Replacing the operators $a_x$
by their Fourier transforms yields
\beq
K_{eff} & = & \frac{1}{4 N^2}  \sum_{k,k',\nu } \la a_{k'}^{\d} a_k \ra
(e^{ik \nu} + e^{-ik' \nu} ) \sum_x e^{ i(k -k') x } \nonumber \\
&  = & -  \frac{1}{4 N} \sum_k \la a_{k}^{\d} a_k \ra \eps_k
\eeq
Finally
\be
K_{eff} = - \frac{1}{4 N} \sum_k n(k)  \eps_k .
\ee
The last equation is very convenient since the mean field formula for $n(k)$
has been already given.  In \fc we compare the results from simulations and
the mean-field prediction. The observed reduction is smaller than
given by the theory. It is not surprising since the mean field approximation
is known to {\em overestimate} effect of $U$. Indeed the mean field ground
state does not take into account the suppression of double occupancy.
Consequently the effect of the interaction term is overestimated.

\section{Summary and conclusions}

Motivated by the interest for numerical approach proposed by L\"uscher 
we have developed a  Monte Carlo algorithm for studying the two
dimensional  Hubbard Model. Our method is based on the field
theoretical formulation in $(2+1)$ dimensions, to which we applied the
polynomial approximation proposed by L\"uscher \cite{lu}. Finally we
obtained the local action and therefore we were able to study the
system using bosonic techniques. We have introduced a number of
improvements which reduce the long autocorrelation times. In
particular we show that by introducing a preconditioning of the
fermionic speeds up the algorithm significantly. We have also
compared the efficiency of the  L\"uscher algorithm with the simple
algorithm based on the direct update of  the determinant of the
fermionic matrix. It follows from this comparison that at present
stage L\"uscher technique does not seem to be a true alternative to
the standard simulations  of the Hubbard Model. The L\"uscher method
can not reach the most interesting region of strong  coupling because
then the fermionic matrices are very badly conditioned,  much more
worse than in the case of QCD. A large number of the  additional
bosonic fields is required and the algorithm suffers from  strong
autocorrelations. Another difficulty arises from nonhermicity 
of the fermionic matrix of the Hubbard Model which complicates the resulting 
action and squares the condition numbers for matrices entering the 
problem. The nonhermitian version of the L\"uscher algorithm can not be 
also applied.

The algorithm based on the direct update of the determinant of the fermionic
matrix was modeled after the algortihm from Ref. \cite{white}.
However we found that numerical instabilities do not appear in
our particular representation of path integrals, which allow to avoid 
additional complications. Hence we managed to reach region of strong 
coupling 
and quite large $\beta$ at half-filling with the help of this very simple 
algorithm. Whenever it was possible we also compared our results with results
of previous investigations as well with the mean field theory.
 
To show the that the ground state has antiferromagnetic long range order
we measured the correlation between two different sites on the lattice.
We found that main limitations in measurements of this quantity are 
finite $\nt$ effects which introduce significant bias. Nevertheless, 
through the extrapolation $\nt \rightarrow \infty$, we 
were able to obtain results which are in a qualitative agreement with 
other QMC simulations  \cite{white} \cite{scalapino}. We have also attempted 
to measure the one-particle 
properties. In particular we showed that the Fermi surface is not sharp
with the shape  described well by the mean-field  theory.  This shows that 
the system is in the insulating Neel state. Thus we did not found any 
unexpected behavior and our results confirm the often presented opinion 
that the physics of the half-filled Hubbard Model is well understood.

On the other hand there is the case of non half-filled band with possible
existence of the superconducting phase. Although the
Projector Quantum Monte Carlo simulations delivered substantial progress
in this field the situation still remains unclear. Especially  just
below the half-filling where the sign problem is a serious obstacle. 
Current simulations are consistent with the most orthodox point of view
treating holes as quasiparticles but other non Fermi liquid
scenarios are not completely excluded \cite{dag}  
\cite{scalapino} \cite{scal95}. It is also impossible to make final 
conclusions about existence of the superconducting phase because 
the range of parameters accessible to simulations is too narrow.

\begin{center}
{\large Acknowledgments}
\end{center}

I would like to specially thank my thesis advisor J.~Wosiek for the
continuous interest in my work, helpful comments and discussions.
I have benefited also from discussions with Z.~Burda and K.~Ro\'sciszewski.
I am very grateful to the Foundation for Polish Science for the fellowship.
The work was partially supported  by the KBN grants no.~2P03B19609
and no.~2P03B04412. The nuumerical calculations were partly performed 
at the ACK Cyfronet-Krak\'ow. Computational grant: KBN/UJ/055/94.

\newpage
\appendix
\section{Appendix A}
In this Appendix we apply the mean field theory \cite{mean} to the
half-filled Hubbard Model on the square lattice with the antiferromagnetic
order in the ground state. Within this approximation one rewrites the
operator of the number of electrons at a given site $i$ as  $ n_{i
\s} =\la n_{i \s} \ra +( n_{i \s} - \la n_{i \s} \ra ) $ where $ \la
n_{i \s} \ra $ is the expectation value in the ground state.
Then assuming the fluctuations to be small, the interaction term in
Hamiltonian becomes
\be
\niu \nid = - \la \niu \ra \la \nid \ra +
\niu \la \nid \ra + \nid \la \niu \ra
\ee
It is convenient to choose as the ground state Spin Density Wave
(SDW) state defined by
\be
\la \niu \ra = \frac{1}{2} \left[ 1 + m(-1)^i \right], \; \;
\la \nid \ra =
\frac{1}{2} \left[ 1 - m(-1)^i \right]
\ee
where as usual $(-1)^i \equiv (-1)^{i_x +i_y}$ is the parity of site,
and $m$ is a variational parameter. Its value will be determined from
the minimum energy condition. Note that  we are able to reproduce the
result for interacting electrons putting $m=0$ and the
antiferromagnetic Neel state $(m=1)$. Transformation of the original
Hamiltonian to the momentum space yields the
mean field Hamiltonian in the form
\be
{\cal H}_{MF}= \sum_{ {k} \s} \eps_{k} a_{k \s}^\dagger a_{k \s}
- \frac{U m }{2} \sum_k (a^{\dagger}_{p+Q \uparrow} a_{k \uparrow} +
a^{\dagger}_{p+Q} a_{k \downarrow} ) + N \frac{U m^2}{4} .
\ee
where  $\eps_k = -2K(\cos{ k_x} + \cos{k_y} ) $ are the bare energy levels
and $Q=(\pi , \pi)$ .
The above Hamiltonian is diagonal in the spin indices and the operators
with momentum $k$ can interact only with those of momentum $k+Q$. Thus
our problem reduces to the diagonalization of the set of matrices
\be
\left(
\begin{array}{cc}
\eps_k   & \pm mU/2 \\
\pm mU/2   &  \eps_{k+Q}
\end{array}
\right)  .
\ee
The eigenvectors and eigenvalues can now be simply obtained
\beq
\gamma_{k \uparrow}^{(+)} = u_k a_{k \uparrow} - v_k a_{ k+ Q \uparrow}, \; \;
\gamma_{k \downarrow}^{(+)} = u_k a_{k \downarrow} + v_k a_{ k+ Q
\downarrow},
\nonumber \\
\gamma_{k \uparrow}^{(-)} = v_k a_{k \uparrow} + u_k a_{ k+ Q \uparrow}, \; \;
\gamma_{k \downarrow}^{(-)} = v_k a_{k \downarrow} - u_k a_{ k+ Q \downarrow}
.
\eeq
There are two eigenvectors per spin and momentum  $k$ is in the
reduced Brillouin zone. The transformation amplitudes $u_k$ and $v_k$ are
\beq
u_{k}^{2} = \frac{1}{2} \left ( 1 + \frac{\epsilon_k}{ E_k} \right ), \nonumber
\\
v_{k}^{2} = \frac{1}{2} \left ( 1 - \frac{\epsilon_k}{ E_k} \right ), \nonumber
\\
E_k = {(\eps_{k}^{2} + \Delta^2 )}^ {1/2}, \nonumber \\
\Delta = - \frac{m U}{2 } . \label{ampl}
\eeq
where $\Delta$ is the energy gap parameter and $ \pm E_k$
are the eigenvalues of $\gamma^{(\pm) }_{k}$. At the edge of the Fermi
surface $k=(-\pi /2, \pi /2)$ we have two possible energies $E= \pm \Delta$
thus the system acquires gap equal $2 \Delta$. The mean field ground state
is obtained by filling the states with lower energy (i.e. those
with the sign (-) in front of $E_k$).
\be
| \psi_{MF} \ra = \prod_{|k| \leq k_{F} }
{ (\gamma_{k \uparrow}^{(-)}) }^\d
{(\gamma_{k \downarrow}^{(-)}) }^\d
|0 \ra ,
\ee
with the empty state denoted by $|0 \ra $.
The condition for energy minimum leads to the self-consistent equation for
the energy gap
\be
\frac{1}{U} = \frac{1}{N} \sum_{|k|  \leq |k_F| }
\frac{1}{ {(\epsilon_{k}^{2} + \Delta^2)}^{1/2} }
\ee
where the sum over momenta is restricted to half of Brillouin zone
($k_F$ is the momentum at the noninteracting Fermi surface).
The above equation for $\Delta$ for finite lattices can be easily
solved numerically. The mean field predictions can be simply obtained
for physically interesting quantities. For example the double
occupancy of the site
is simply related to the antiferromagnetic order parameter $m$
\be
\la \niu \nid \ra = \la \niu \ra \la \nid \ra =\frac{1}{4}(1 -m^2).
\ee
One can also obtain the occupation number in the momentum
space
\be
n_{k \s} = \la \psi_{MF}| a^{\d}_{k \s} a_{k \s} | \psi_{MF} \ra
=\frac{1}{2} \left( 1 - \frac{\eps_k}{E_k} \right). \label{ocup}
\ee
Indeed the electrons with momentum $k$ are created from the state $|0 \ra $
with the amplitude $v_k$, given by Eq. (\ref{ampl}) if the momentum $k$ lies
inside the Fermi surface. Similarly the electrons outside the Fermi
surface are created with the amplitude $u_{k -Q}= v_k$.

\section{ Appendix B}

\newcommand{\psis}{\psi^\star }
\newcommand{\psii}{\psi^{ ' \star} }
In this Appendix we review the formalism of Grassmann variables
\cite{creutz} \cite{creutz1}.
A set of anticommuting Grassmann variables is defined by the relation
\be
{[ \psi_i , \psi_j ] }_+ \equiv \psi_i \psi_j  + \psi_j \psi_i =0 .
\ee
For each $\psi_i$ there is also corresponding independent
variable $\psi_{i}^{\star}$. Moreover we have
\beq
(\psi_{i}^{\star})^{\star} & = & \psi_i \; \; \; , \nonumber \\
(\psi_{1} \cdots \psi_{n} )^\star & = & \psi_{n}^{\star}
\cdots \psi_{1}^{\star} \; \; \; .
\eeq
Assuming the properties of linearity and invariance under translation
of variables the integration can be defined through the formulas
\beq
\int d\psi \, \psi =i \nonumber \\
\int d \psi \, 1 =0 \nonumber \\
\int d \psi^\star \, d \psi \, \psi^\star \, \psi =1 .
\eeq
Since the general function of Grassmann variables is in fact
polynomial the above equations are sufficient to compute
any integral. In addition to integration we consider the
differentiation with the respect to the Grassmann variable.
This can be defined by the action on a constant function and an
anticommutation relation
\be
\frac{d}{d \psi} 1 = 0 \; \; \;  , \left[  \frac{ d}{ d \psi^\star}, \psi^\star
\right]_+ =1.
\ee
Function of Grassmann variables form the Hilbert space on which transfer
matrix acts. To make this statement more specific we consider as an example
the case of one degree of freedom. Then one can relate the function
$f(\psi^\star) =1 $ to the vacuum state $|0 \ra$ and
the function $f(\psi^\star) = \psi^\star$ to the occupied state
$|1 \ra$. The  operators  $a^\dagger$ and $a$ would correspond to
$\psi^\star$ and $\frac{d}{d \psi^\star}$ respectively.

It is convenient to introduce the simple analog of Dirac $\delta$ function
for the anticommuting variables
\be
\delta (\psi^\star , \psi^{ ' \star} ) =
\int (d \psi) e^{ \sum ( \psi^\star -\psi^{ ' \star} ) \psi } ,
\ee
where $(d \psi)$ denotes $\prod d \psi_i $ in some prescribed order.
Note that the $\delta$ function has the expected property
\be
f( \psis ) = \int \delta (\psis , \psii ) {(d \psi^{' })}^\star f( \psii) .
\ee
Now it is easy to verify the  correspondence on the level
of integral representations for operators acting on a general
state $|f \ra $
\beq
a_i |f \ra  \leftrightarrow \int \psi_i \, (d \psi)
e^{\sum (\psi^\star -\psi^{ ' \star}) \psi } \, {(d \psi^{'} )}^\star
f(\psi^{ ' \star} ) \;, \nonumber \\
a_{i}^{\dagger} |f \ra  \leftrightarrow \int \psi_{i}^{\star} \, (d \psi)
e^{\sum (\psi^\star -\psi^{ ' \star}) \psi } \, {(d \psi^{'} )}^\star
f(\psi^{ ' \star} ) \;  .
\eeq
The result is very simple and states that under integral
$a$ and $a^\dagger$ should be replaced by $\psi$ and $\psi^\star$
respectively. This immediately generalizes to any normal ordered
function of $a$ and $a^\dagger$
\be
:g(a^\dagger, a): |f \ra
\leftrightarrow \int g(\psi^\star , \psi) \, (d \psi)
e^{\sum (\psi^\star -\psi^{ ' \star}) \psi } \, {(d \psi^{'} )}^\star
f(\psi^{ ' \star} )  \nonumber \\
.
\ee
It is very important that the function $g$ is normal ordered. Variables
$\psi$ and $\psis$ simply anticommute while the anticommutation relation
between $\psis$ and $\frac{d}{ d \psis}$ introduces additional factors.

The trace of a normal ordered operator can be rewritten in the form of
integral over anticommuting variables
\be
Tr \; :g(a^\d, a): \; = \int (d \psi)^\star g(\psis, \psi) e^{ 2 \sum \psis
\psi } (d \psi )  .
\ee
A little algebra  yields the trace of normal ordered factors as a multiple
integral over anticommuting variables
\beq
Tr [:g_1(a^\dagger, a)::g_2(a^\dagger, a): \cdots : g_{\nt} (a^\dagger, a):]=
\nonumber \\
\int \prod_{t=1}^{\nt} [ {(d \psi_t)}^\star g_t (\psi_{t}^{\star}, \psi_t)
(d \psi_t) e^{ \sum \psi_{t}^{\star} (\psi_t -\psi_{t-1}) } ] ,
\eeq
with antiperiodic boundary conditions $ \psi_0= -\psi_{\nt}$.
For operators $g$ being the transfer matrices of the Hubbard Model
the integration is performed by the standard formula
\be
\int {(d \psi)}^\star \, (d \psi) e^{ \psi^\star M \psi}= \det{M}.
\ee
which directly gives Eq. (\ref{grassman}).

The expectation value of observables just inserts another factor in the
above integral. The additional factors under integral are
\be
\int {(d \psi_{\nt +1})}^{\star} \, (d \psi_{\nt +1}) \,
h(\psi_{\nt +1}^\star, \psi_{\nt +1}) \exp{ [ \psi_{\nt +1}^{\star}
(\psi_{\nt +1} - \psi_{\nt} ) + \psi_{1}^{\star} (\psi_{\nt +1} -\psi_{\nt} )
]} .
\ee

As an example we consider the expectation value for the numbers of electrons
with spin up and the corresponding function $ :h(a^\dagger, a): =a^\dagger a$.
Then the integration over additional factors reduces simply to
the insertion $ 1- \psi_{1}^{\star} \psi_{\nt} $ under the main integral.
The evaluation of integral of such type is simple if we make use of the
identity
\be
\int (d \psi)^\star d \psi e^{\psis M \psi + \xi \psi + \psis \eta  } =
\det{M} \;  e^{ - \xi M^{-1} \eta } .  \label{int}
\ee
The additional factors can be obtained simply by differentiation with
the respect to the anticommuting variables $\eta$ and $\xi$ next
putting them zero. In our example
\be
\int {(d \psi)}^\star \, (d \psi) e^{ \psi^\star M \psi}
 \; \; (1- \psi_{i,1}^{\star} \psi_{i,\nt})  = \det{M} \; \; \;
 ( 1 - M_{i,\nt; i,1}^{-1} ).
\ee
Cyclically shifting in $t$ direction yields
\be
\la n_{i \uparrow} \ra = \la 1 + M_{i,t; i,t +1}^{-1} \ra.
\ee
In similar derivation for the electrons with spin spin down one should
first perform the particle hole transformation $a_{\downarrow}^\d
a_\downarrow = b b^\d = 1 - b^\d b $. The corresponding insertion is
$\psi_{1}^{\star} \psi_{\nt} $ and the final result reads
\be
 \la  n_{i \downarrow} \ra = - \la  M_{i,t; i,t +1}^{-1} \ra .
\ee
Similar considerations give the expressions for the other observables of
interest
\beq
\la n_{i \uparrow} n_{i \downarrow} \ra & = & - \la M_{i,t;i,t+1}^{-1}
                                       (1 + M_{i,t;i,t+1}^{-1}) \ra \\
C(i,j) & \equiv & \la (n_{i \uparrow} - n_{i \downarrow})  
(n_{j \uparrow} - n_{j \downarrow}) \ra \nonumber \\
C(i,j) &  = & \la 1+2 M_{i,t; i,t +1}^{-1} + 2 M_{j,t; j,t +1}^{-1}
 -2 M_{i,t;i,t+1}^{-1} \delta_{i,j}    \nonumber \\
 & & \hspace{2.2cm} + 4 M_{i,t; i,t +1}^{-1} M_{j,t; j,t +1}^{-1} -
 2 M_{i,t; j,t +1}^{-1}  M_{j,t; i,t +1}^{-1} \ra  \\
n(k) &= & \frac{1}{N} \sum_{x y} e^{i k (x -y)} \la M_{x,t;y,t+1}^{-1}
- (-1)^x (-1)^y M_{x,t;y,t+1} + \delta_{ x,y} \ra .
\eeq

\newpage
{\Large {\bf Figure Captions }  }
\begin{enumerate}

\item Fermi surface for the 2d Hubbard Model on the square lattice.
The dashed area corresponds to the occupied states for half-filling.

\item Errors of the L\"uscher approximation measured on a single typical
configuration genrated at $U=1$ and $\beta=1$. The quantity 
$\delta$ defined in text is shown as a function of $\eps$. 
The solid, dashed and dotted line is for number of fields $50$, $70$ 
and $100$ respectively.

\item  The distribution of the condition numbers for the $M^\d M$
matrix (solid line) and the preconditioned matrix $M^\d D^{-1} M$
(dashed line). Results from simulations on $6^2 \times 12$ lattice
at $U=4$ and $\beta=1$.

\item Eigenvalues of the matrix $M$ plotted on the complex plane.
A typical configuration was taken from simulations at $U=1$ and $\beta=1$. 

\item The double occupancy versus U on a $4 \times 4$ lattice
with $\beta=5$. The solid line is the mean-field result.

\item
Magnetic correlation function $C(l)$ on the $4 \times 4$ lattice.
The point $l$ traces the  triangular path shown in the picture.
The antiferromagnetic correlations are clearly visible at $\nt=60$,
$\beta=5$ and $U=4$. The shape of curve is very sensitive to the finite 
$\nt$ effects. The results at $\nt=30$ (dotted line) and $\nt=40$ 
(dashed line) are also shown for comparison.

\item
Magnetic factor $S(k_x, k_x)$ along the main diagonal of the
Brillouin zone. Results from lattices $6 \times 6$ (empty circles), 
$8 \times 8$ (triangles), $12 \times 12$ (circles) and 
$16 \times 16$ (crosses) are collected. Other parameters of runs were $U=1$
and $\beta =1$. There is no signal of long range antiferromagnetic order
at this temperature. 

\item
Finite $\nt$ analysis of antiferromagnetic factor $S(\pi, \pi)$.
Its value is extrapolated in the parameter ${ (\frac{\beta}{\nt}) }^{2}$.
Results from lattices $4 \times 4$ (circles) and $6 \times 6$ 
(triangles) are presented. The extrapolation with $\nt \rightarrow \infty$
gives result $2.6 \pm 0.2$, and $3.2 \pm 1.0$ respectively.

\item
The ferromagnetic factor $S(0,0)$ as a function of spatial  volume of the
lattice. Three sets of data are shown for $\nt=6$ (circles), $\nt=8$
(triangles), and $\nt=12$ (crosses).

\item
The distribution of electrons $ \la n_{k \uparrow}+n_{k \downarrow} \ra $
in the momentum space. Monte Carlo results from lattice $6 \times 6$ 
(crosses) and $8 \times 8$
(triangles) are shown. They are well described by the mean-field theory 
(solid line). Dashed line is a noninteracting Fermi distribution 
$f(\eps_k)= 2 (\exp(- \beta \eps_k) + 1)^{-1}$.

\item
The effective hopping versus U on a $4 \times 4$ lattice with
$\beta =5$. The solid line is the mean field result.

\end{enumerate}

\newpage

\begin{figure}
\includegraphics{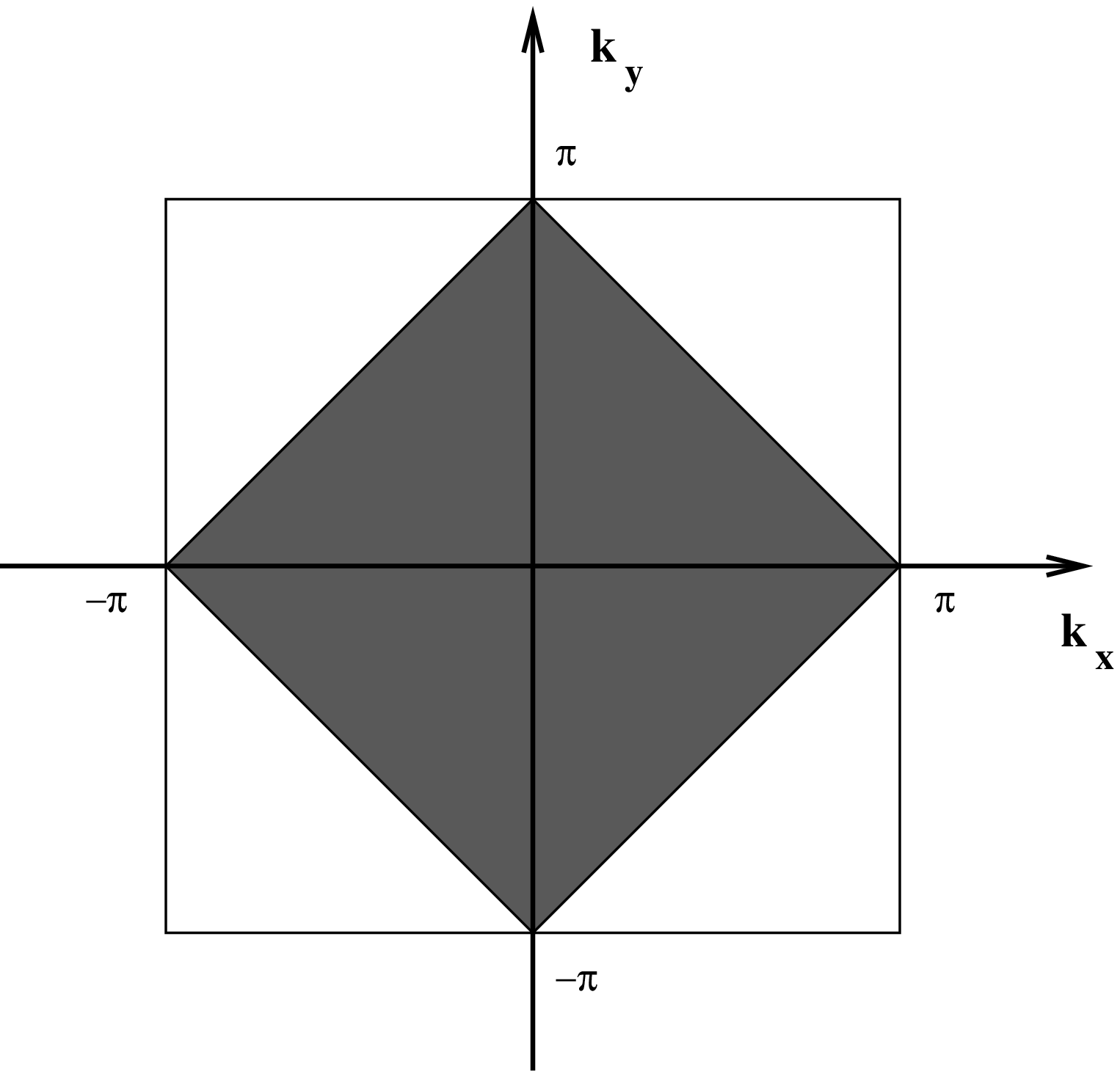}
\vspace{2cm}

\caption{ }
\end{figure}

\newpage

\begin{figure}
\psfrag{delta}{ {\Large $\delta$} }
\psfrag{eps}{ {\Large $\epsilon$ }}
\includegraphics{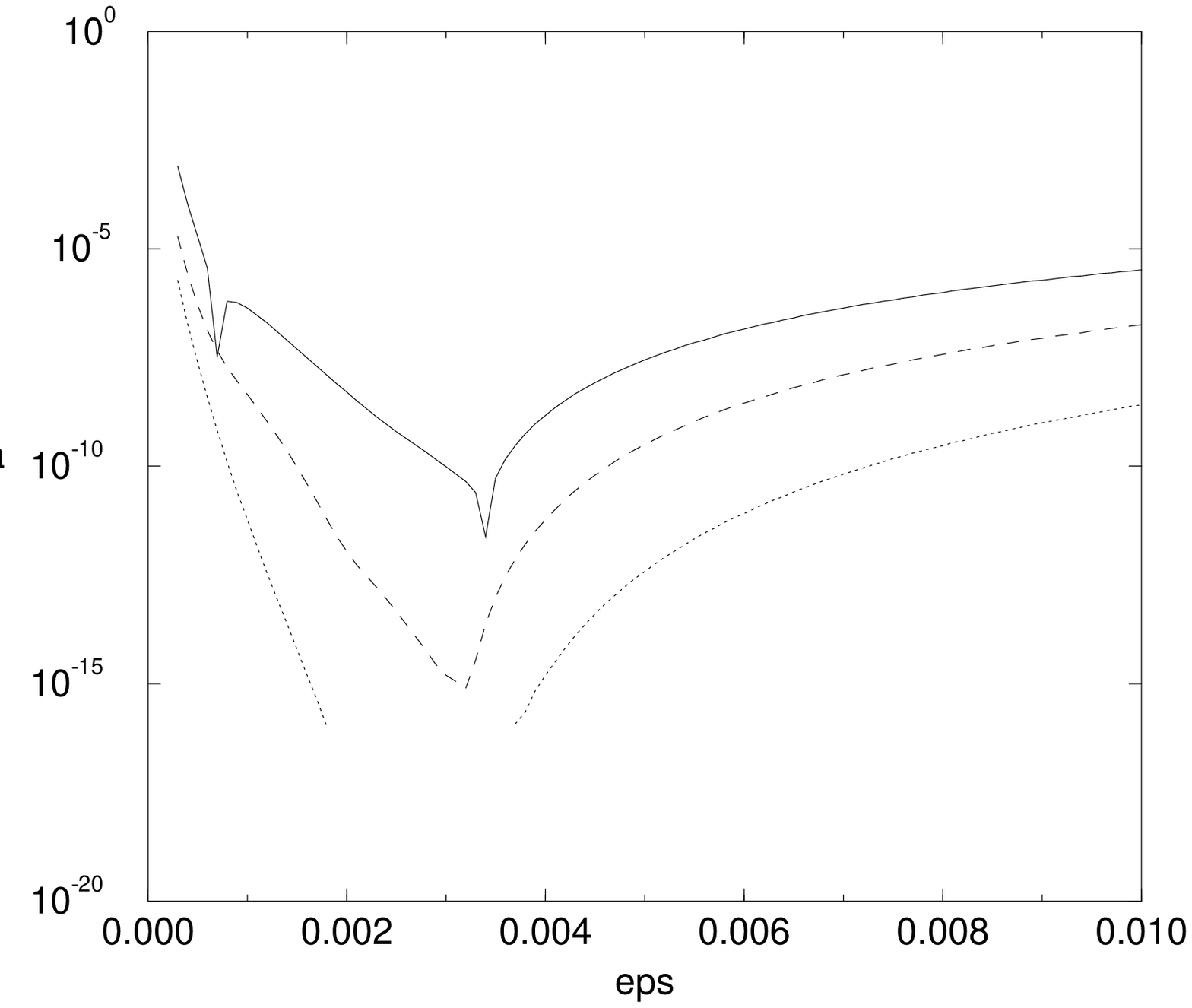}
\vspace{2cm}

\caption{ }
\end{figure}
\newpage

\begin{figure}
\includegraphics{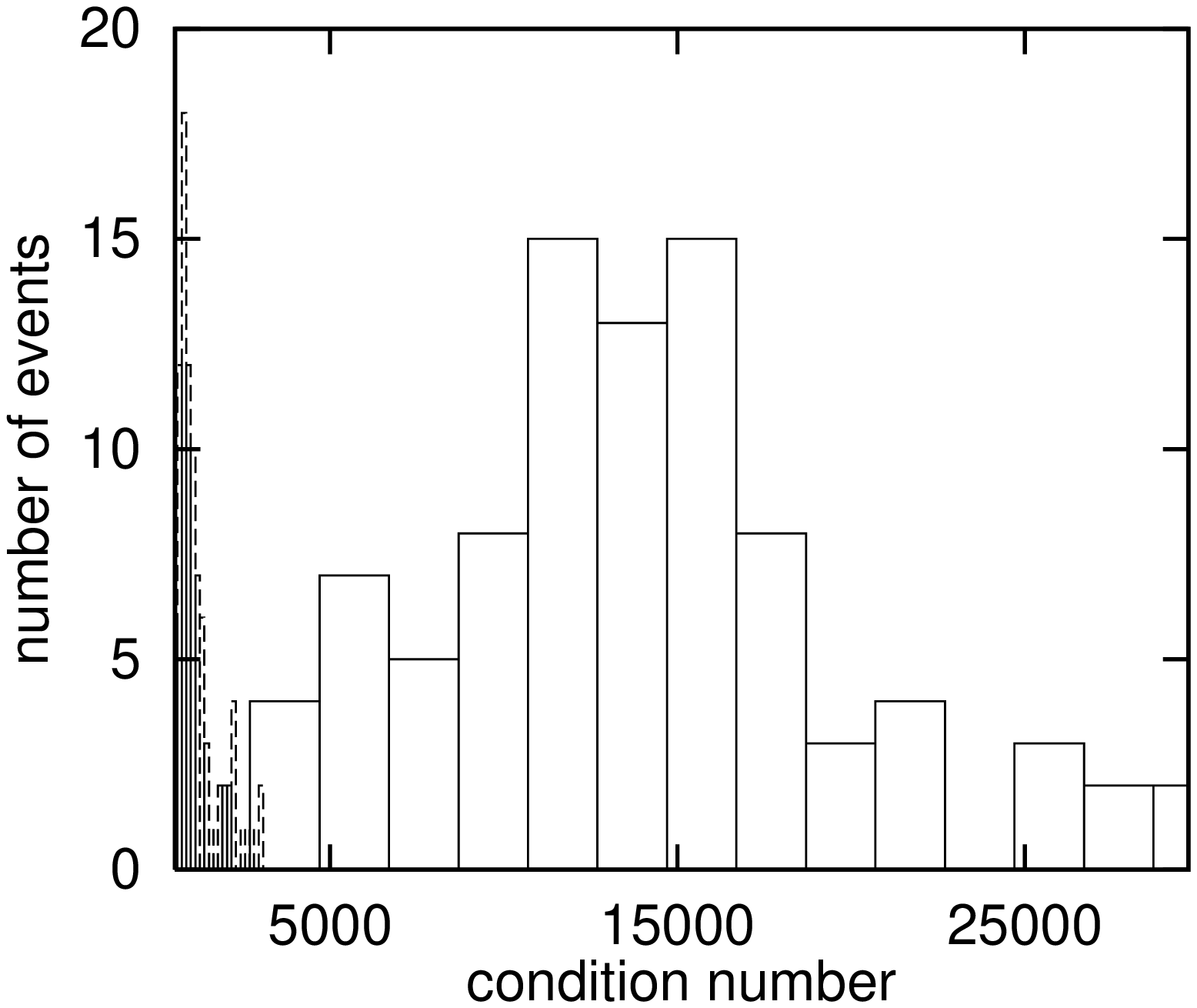}
\vspace{2cm}

\caption{ }
\end{figure}

\newpage

\begin{figure}
\includegraphics{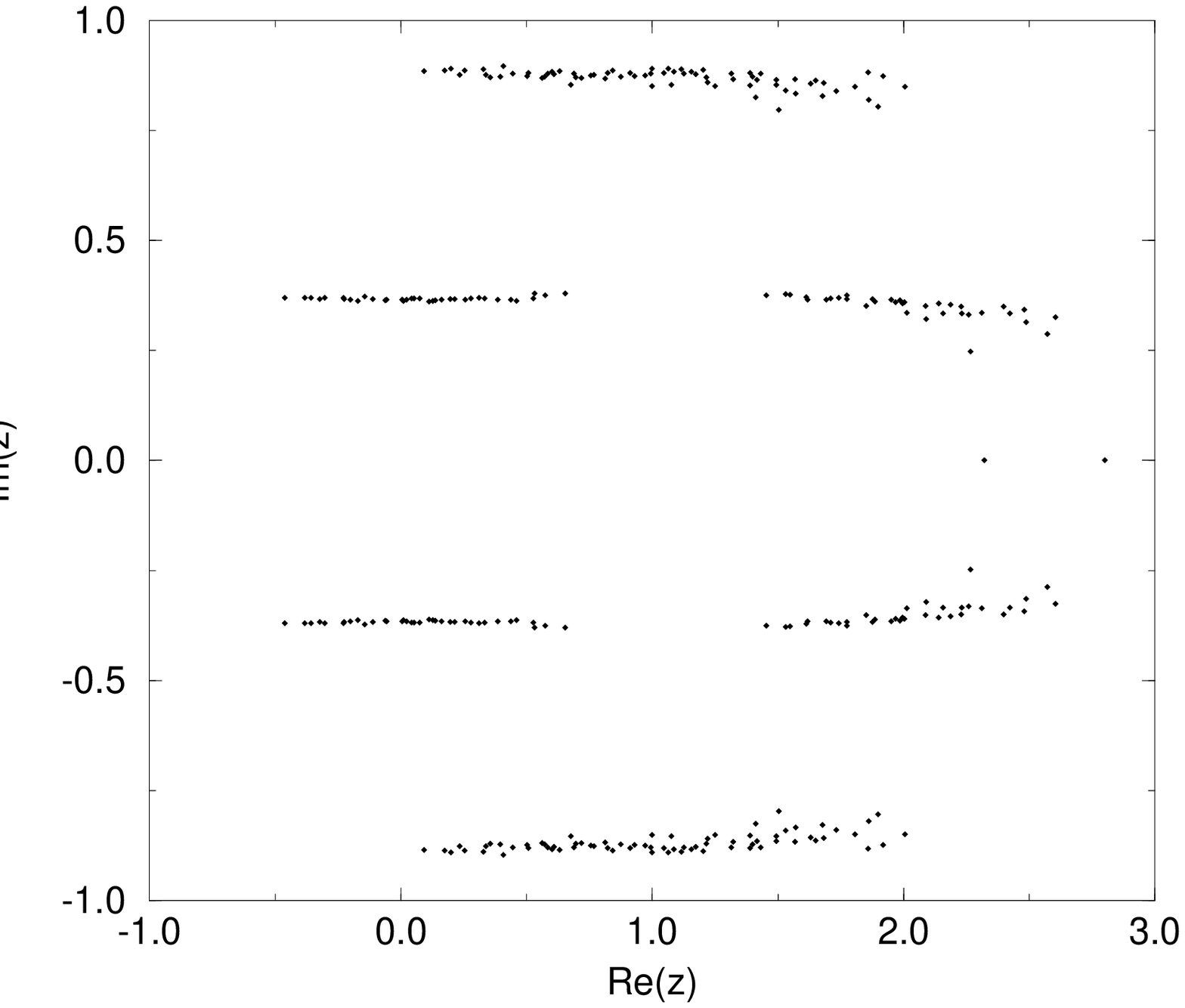}
\vspace{2cm}
\caption{ }
\end{figure}

\newpage

\begin{figure}
\psfrag{pary}{{\Large $\la n_{i \uparrow} n_{i \downarrow} \ra$} }
\includegraphics{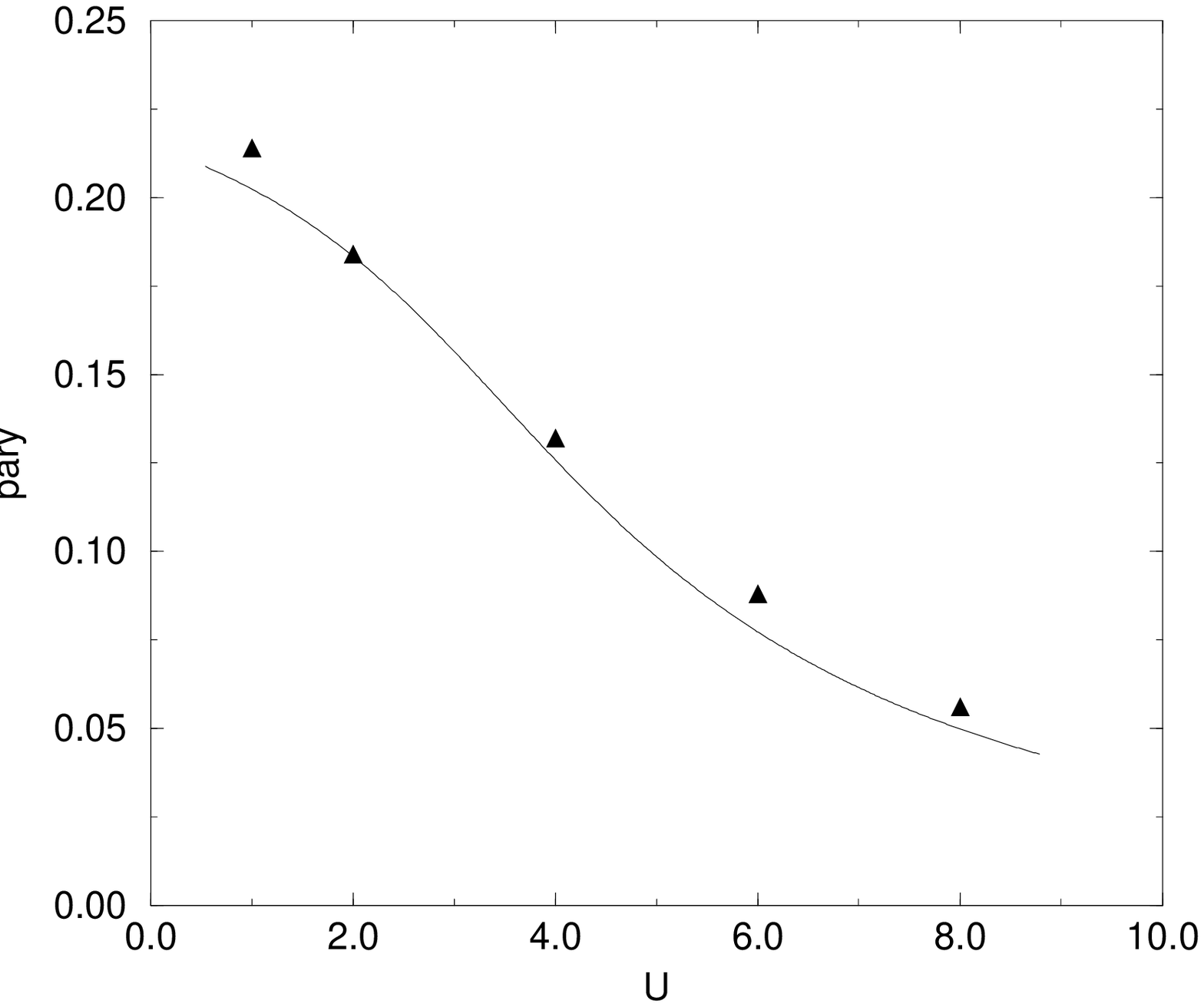}
\vspace{2cm}

\caption{ }
\end{figure}

\newpage

\begin{figure}
\includegraphics{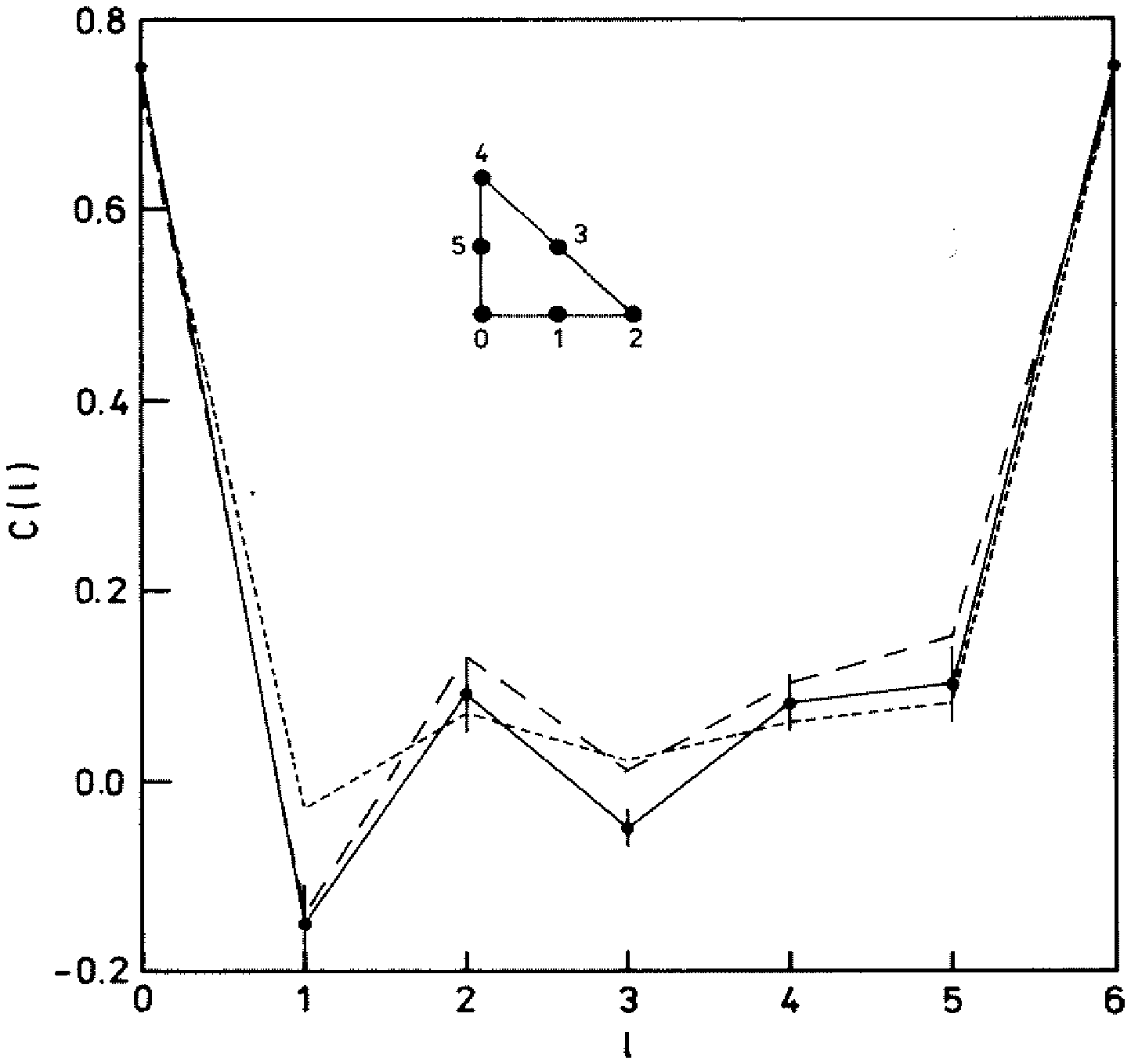}
\vspace{2cm}

\caption{ }
\end{figure}
\newpage

\begin{figure}
\includegraphics{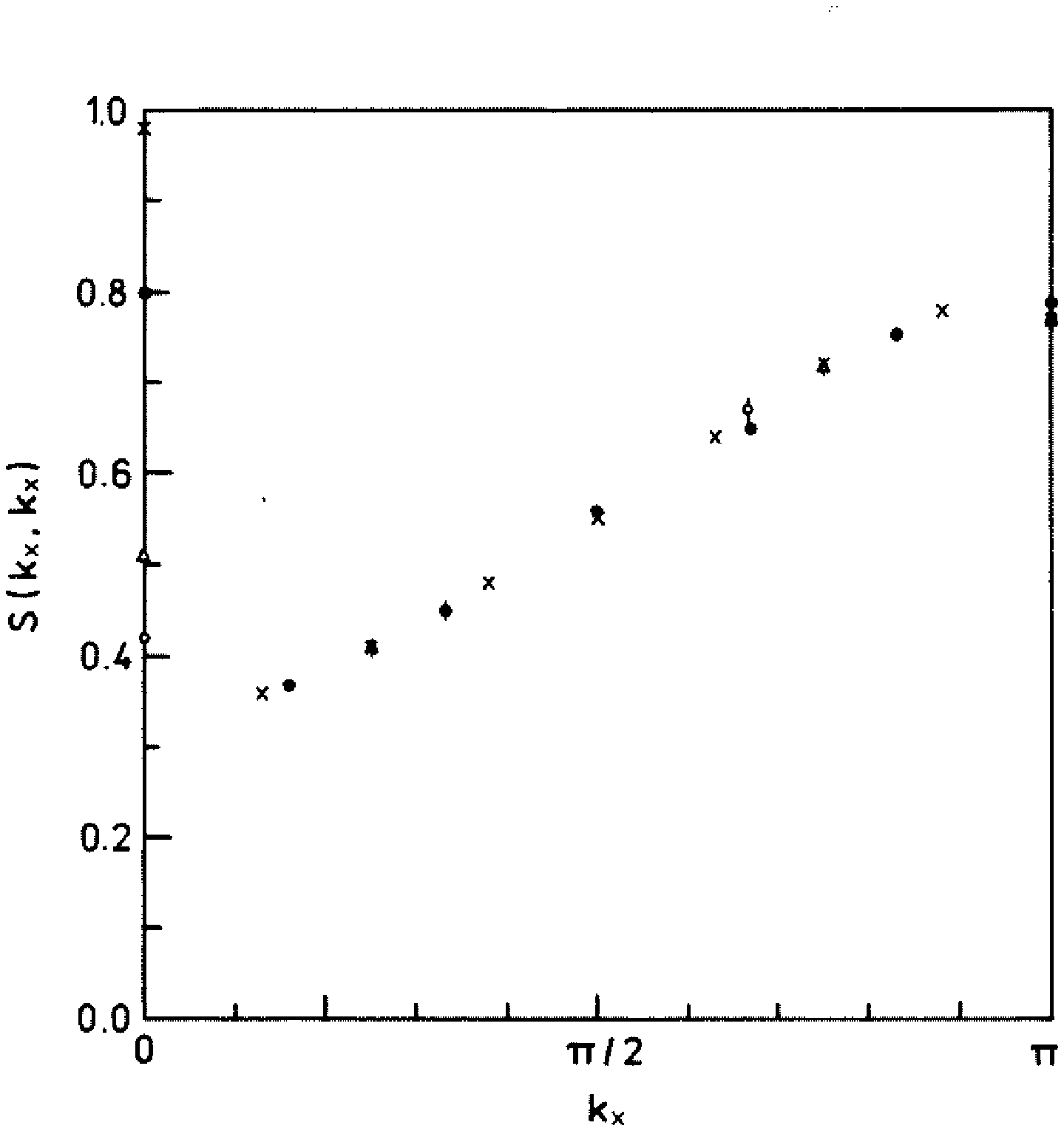}

\vspace{2cm}
\caption{ }
\end{figure}

\newpage

\begin{figure}
\includegraphics{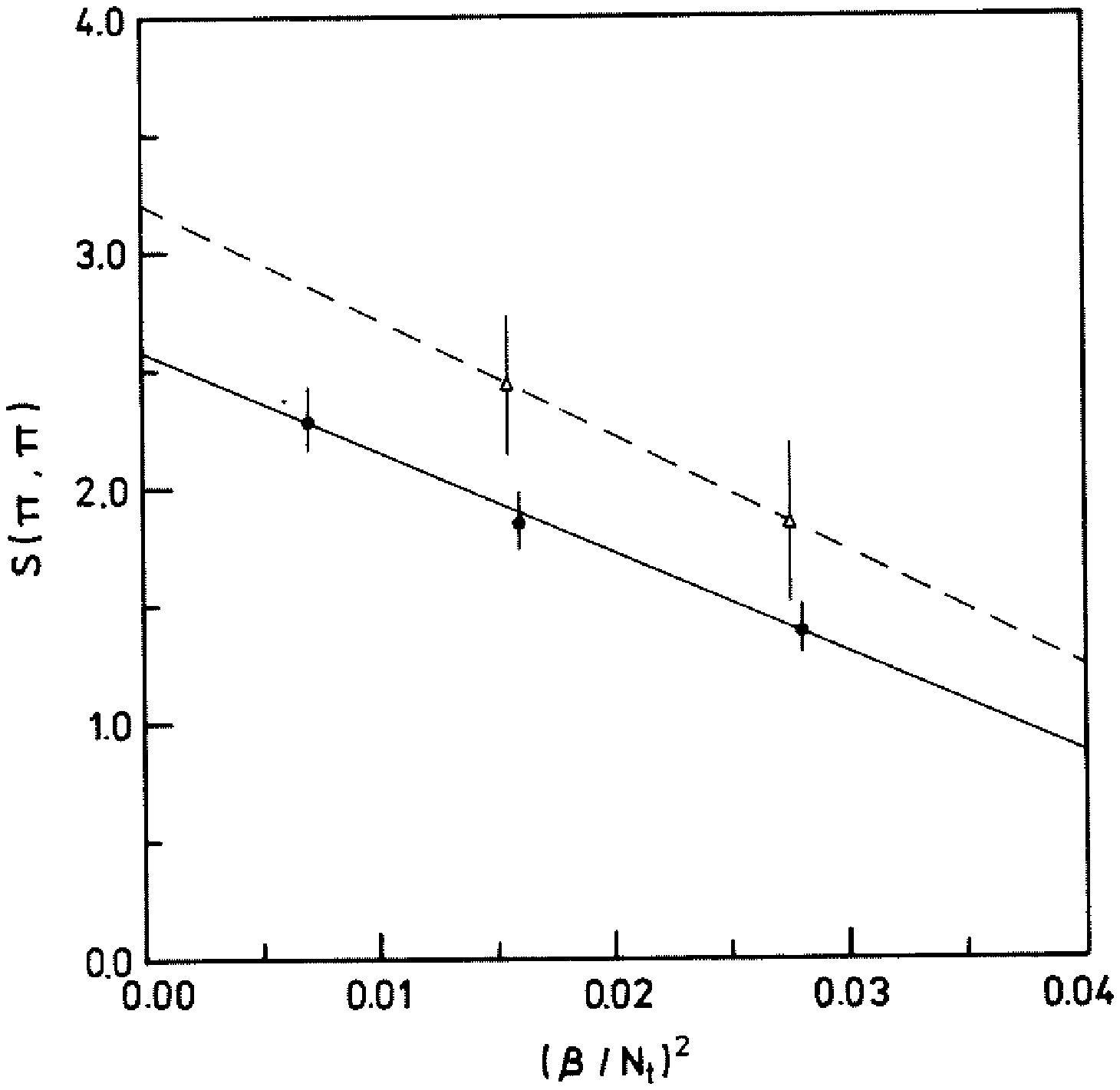}
\vspace{2cm}

\caption{ }
\end{figure}

\newpage

\begin{figure}
\includegraphics{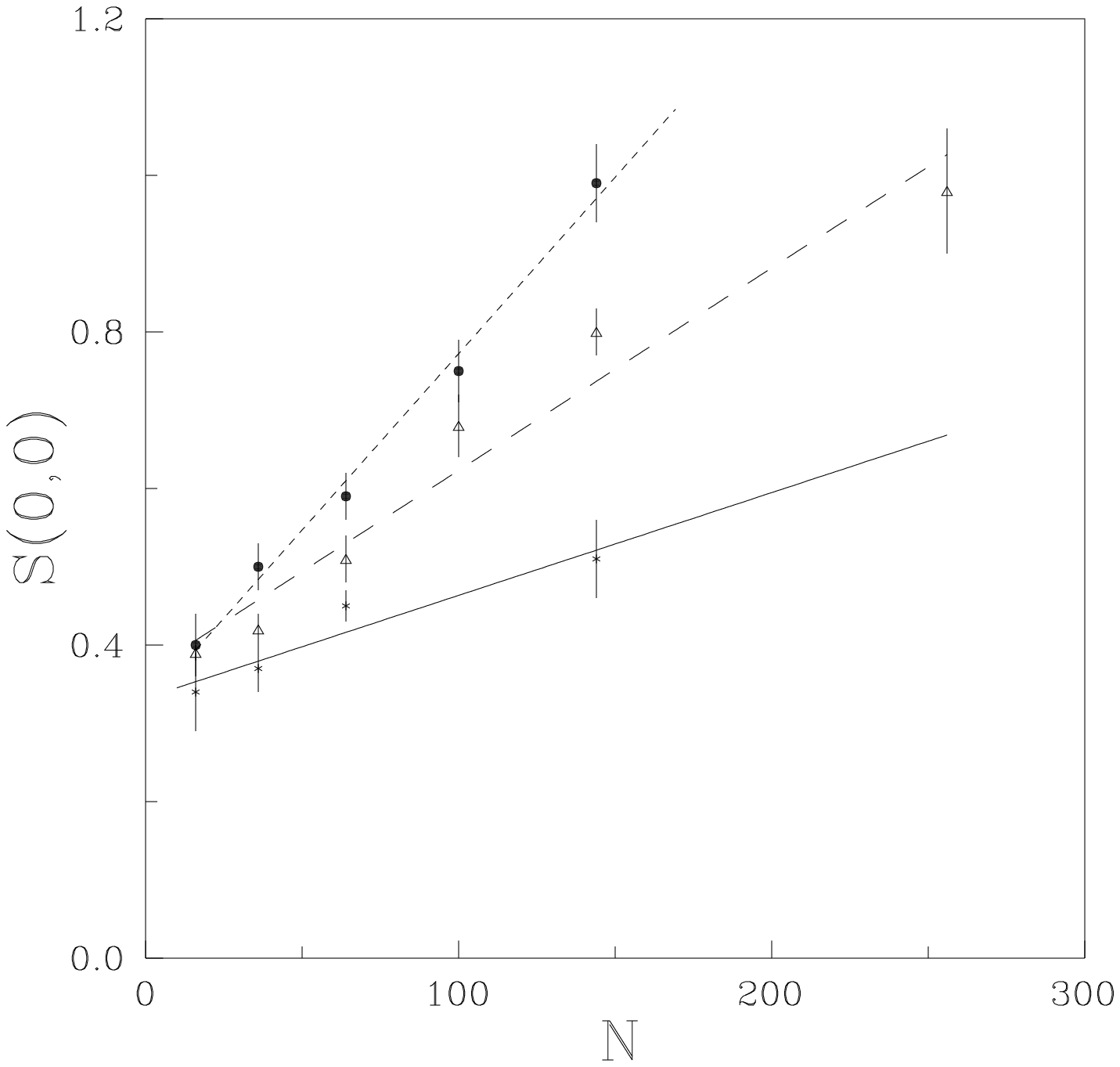}
\vspace{2cm}

\caption{ }
\end{figure}
\newpage

\begin{figure}
\includegraphics{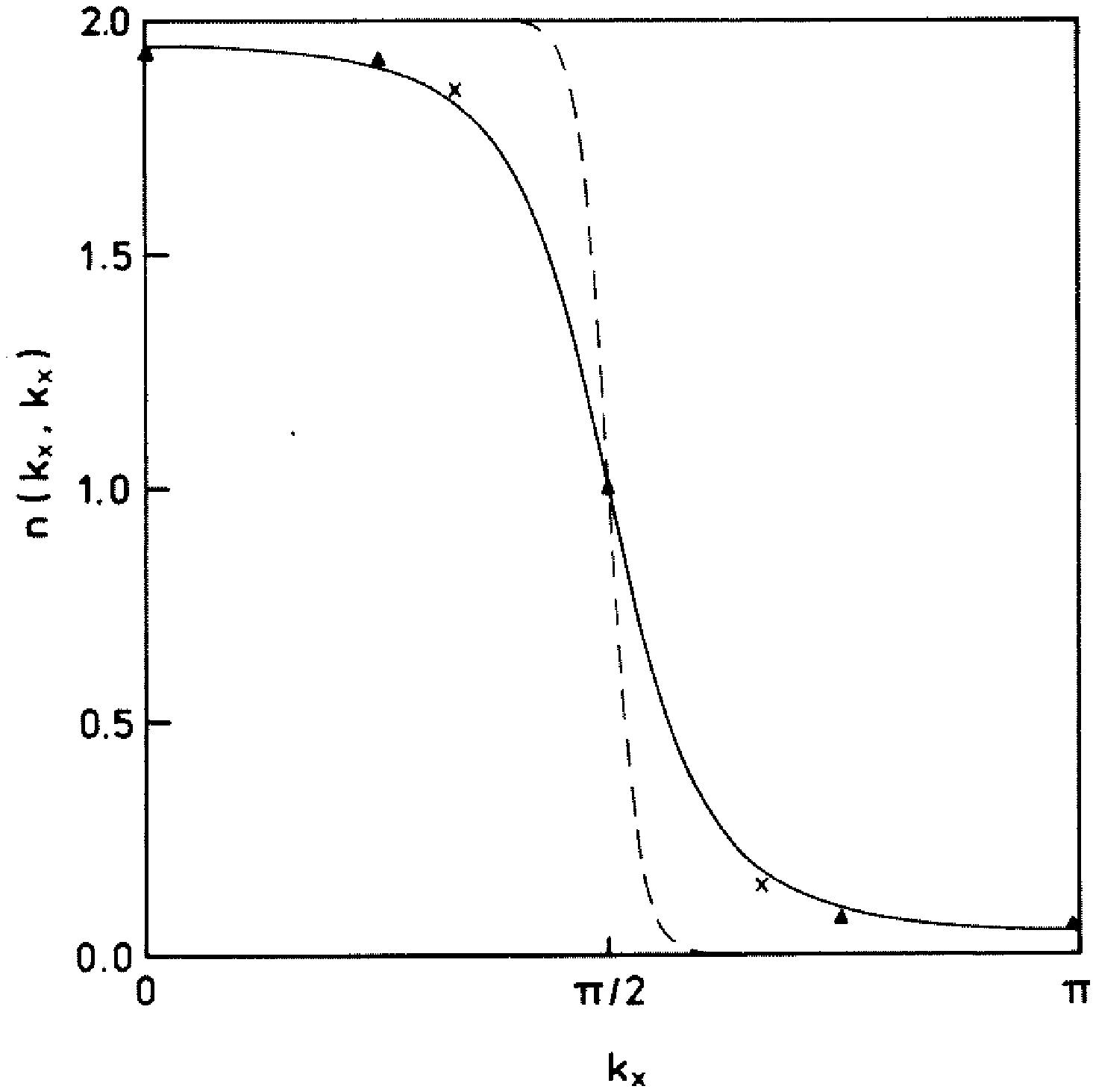}
\vspace{2cm}

\caption{ }
\end{figure}
\newpage

\begin{figure}
\psfrag{hopping}{ {\Large $K_{eff}/K_0$} }
\includegraphics{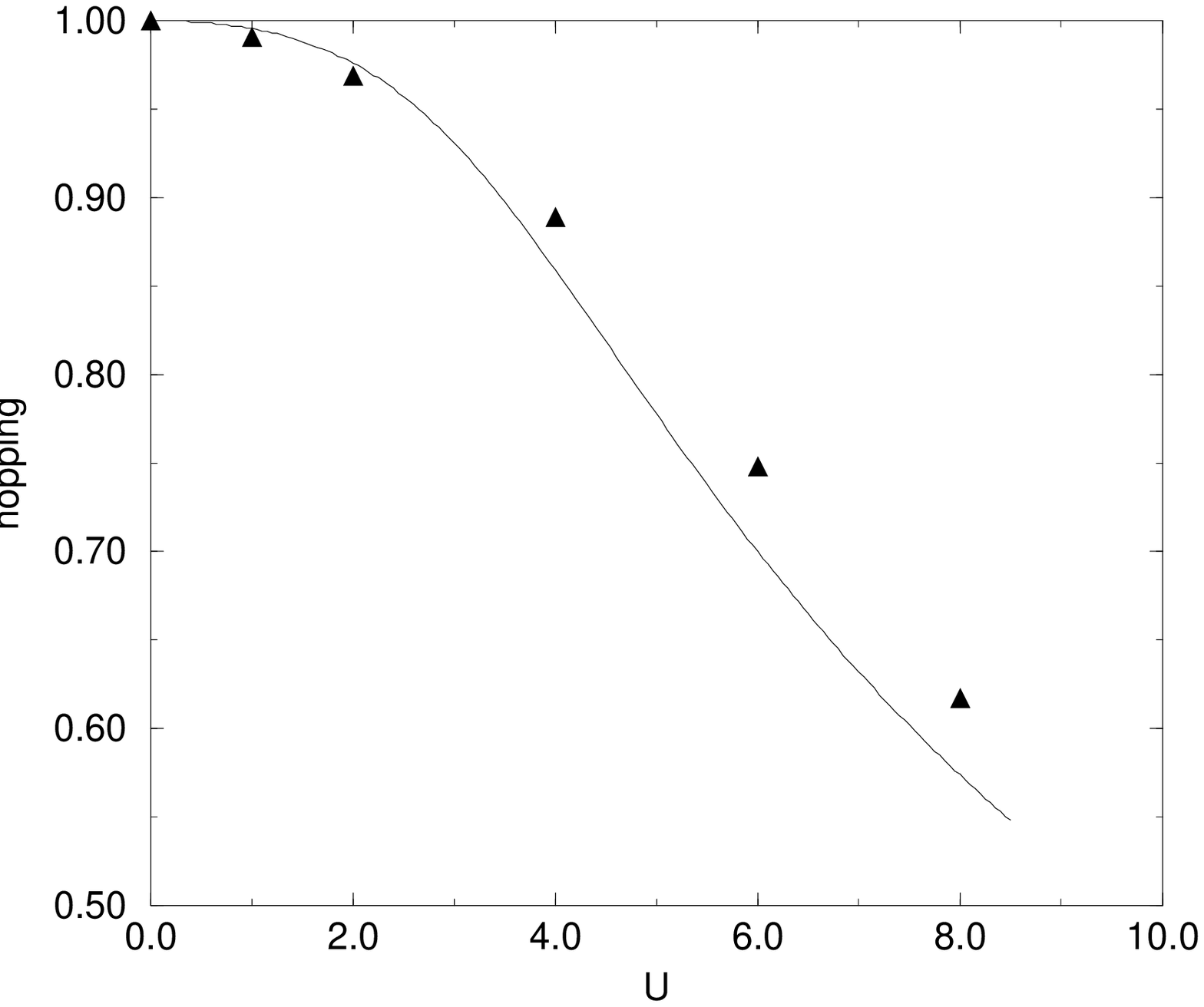}
\vspace{2cm}

\caption{ }
\end{figure}


\newpage
\begin{thebibliography}{99}
\bibitem{qg}  M. Creutz, { \em "Quarks, Gluons and Lattices"},
              Cambridge University Press 1983.
\bibitem{lu} M. L\"uscher, Nucl. Phys. {\bf B418} (1994) 637.
\bibitem{hub} J. Hubbard, Proc. Roy. Soc. {\bf A 276} (1963) 238.
\bibitem{dag} E. Dagotto, {\em "Correlated Electrons in high
temperature superconductors"}, Rev. of Mod. Phys. {\bf Vol. 66 No. 3}.
\bibitem{and} P. W. Anderson, Science {\bf 234} (1987) 1196.
\bibitem{wu}  E.H.Lieb, F.Y. Wu, \PRL {\bf 20} (1968) 1445.
\bibitem{fra} E. Fradkin, {\em "Field theory of condensed matter systems"},
Addison-Wesley Publishing Company 1991.
\bibitem{varia} H.~Yokoyama and H.~Shiba, J.Phys.Soc.Jpn {\bf 56} (1987)
1490; \\
H.~Yokoyama and H.~Shiba, J.Phys.Soc.Jpn {\bf 56} (1987) 3582.
\bibitem{anti} J. E. Hirsh, S. Tang, \PRL {\bf 62} (1989) 591
\bibitem{fet} A.L.Fetter, J.D.Walecka, {\em "Quantum Theory of Many Particle
Systems"}, McGraw Hill, 1971.
\bibitem{white} S.R.~White et al., \PR {\bf B40} (1989) 506.
\bibitem{scalapino} D. J. Scalapino, {\em "Results from Numerical 
Simulations of the 2D Hubbard Model." } in {\em "High Temperature
Superconductivity Proceedings}, Addison Wesley, 1990. 
\bibitem{sorella} S.Sorella, \PR {\bf B46} (1992) 11670.
\bibitem{scal95} D.J.Scalapino, Phys. Rep. {\bf 250} (1995) 329.
\bibitem{santos} R. dos Santos, \PR {\bf B39} (1989) 7259.
\bibitem{negative} R.T.Scalettar and al., \PRL {\bf 12} (1989) 1407.
\bibitem{creutz} M. Creutz, \PR {\bf B38} (1988) 1228.
\bibitem{creutz1} M. Creutz, \PR {\bf D35} (1987) 1460.
\bibitem{wosiek} J. Wosiek, Acta Phys.\ Pol.\ {\bf B18}
(1987) 519 and references therein.
\bibitem{marinari} F.Fucito, E.Marinari, G.Parisi and  C.Rebbi
Nucl. Phys. {\bf B 180} (1981) 369.
\bibitem{hmc} M.~Creutz, {\em "Algorithms for simulation of fermions"}
in {\em "Quantum Fields on the computer" }, World Scientific, 1992.
\bibitem{langevin} G.G.Batrouni, G.R.Katz, A.S.Kronfeld, G.P.Lepage
and K.G.Wilson, Phys. Rev. {\bf D32} (1985) 2736.
\bibitem{micro} J.Polonyi and H.W.Wyld, Phys. Rev. Lett. {\bf 51}
(1983) 2257; \\
J.Kogut, J.Polonyi, H.W.Wyld and D.K.Sinclair, Phys. Rev.
Lett. {\bf 54} (1983) 1475.
\bibitem{ken} S. Duane, A.D.Kennedy, B.J.Pendelton and D. Roweth ,
Phys. Lett. {\bf B195} (1987) 216.
\bibitem{jan} K.~Jansen and C.~Liu,  Nucl.Phys. {\bf B453} (1995) 375.
\bibitem{petersen} R.~Lacaze, A.~Morel, B.~Petersson and J.~Schroeper, {\em
"An investigation of the 2d attractive Hubbard Model "}, preprint 
BI-TP 96/05.
\bibitem{hirsch} J.E.Hirsch, \PR {\bf B 31} (1985) 4403.
\bibitem{kon} G.Sugiyama, S.E.Koonin Ann. of Phys. {\bf 168} (1986) 1.
\bibitem{car} S.Sorella, S.Baroni, R.Car, M.Parinello,
Europhys. Lett. {\bf 8} (1989) 663;  \\
S.Sorella et al. Int. J. Mod. Phys. {\bf 1} (1988) 993.
\bibitem{sign} A. Galli,  {\em "Suppresion of
the negative sign problem in quantum Monte Carlo" }, hep-lat 9605026.
\bibitem{lanczos} H.Q.Lin, J.E.Gubernatis, Comp. Phys. {\bf Vol. 7}
(1993) 400.
\bibitem{saw} P. Sawicki and J. Wosiek, Nucl. Phys. {\bf B42}
(Proc. Suppl.) (1995) 932; \\
P. Sawicki and J. Wosiek, Nucl. Phys. {\bf B47 } (Proc. Suppl.) (1996) 785.
\bibitem{nr} "Numerical Receipes", Cambridge University Press 1986.
\bibitem{sokal} A. Sokal {\em "Monte Carlo Methods in Statistical Mechanics:
Foundations and New Algorithms"} ,  Cours de Troisieme Cycle de la
Physique en Suisse Romande.
\bibitem{metro} N. Metropolis et al., J. Chem. Phys. {\bf 21}, 1087 (1953).
\bibitem{jeg} B. Jegehrlehner, Nucl. Phys. {\bf B42} (Proc. Supll.)
(1995); \\
K.~Jansen, C.~Liu, B.~Jegehrlehner, {\em "Performance Tests of
the Kramers Equation and Boson Algorithms for simulations of QCD" },
preprint DESY95-243.
\bibitem{for2} A. Borici an Ph. de Forcrand, {\em "Systematic errors of
L\"uscher's fermion method and its extensions"}, preprint IPS-95-13.
\bibitem{aut} N. Madras, A. Sokal, J. Stat. Phys. {\bf 50} (1988) 109.
\bibitem{ed} R. G. Edwards et al. , Nucl. Phys. {\bf B354}  (1991) 289.
\bibitem{monty} I. Montvay, {\em "An algorithm for gluinos on the lattice"},
                preprint DESY 95-192.
\bibitem{for1} C. Alexandrou et al. {\em "Full QCD with the L\"uscher
local bosonic action" } preprint UCY-PHY-95/5.
\bibitem{adler} S. L. Adler, \PR {\bf D37} (1988) 458.
\bibitem{mermin} N.D.Mermin and H.Wagner,  \PRL {\bf 17} (1966) 1133.
\bibitem{aza} A. Auerbach, {\em "Interacting Electrons and Quantum 
Magnetism"}, Springer-Verlag 1994.
\bibitem{huse} D.A.Huse, \PR {\bf B37} (1988) 2380.
\bibitem{galli} Ph. de Forcrand and A. Galli, {\em "Exact Local Bosonic 
Algorithm for full QCD"}, hep-lat 9603011.
\bibitem{mean} J. R. Schreiffer, X. G. Wen and S.C. Zhang, Phys. Rev.
{\bf B39} (1989) 11663.
\end{thebibliography}
\end{document}